\documentclass[11pt,a4paper]{article}
\usepackage{jheppub,tcolorbox}
\usepackage[T1]{fontenc}
\usepackage{tikz}
\usepackage{tikz-cd}
\usepackage{subfigure}
\usetikzlibrary {decorations.pathmorphing}
\usetikzlibrary {decorations.markings}
\usetikzlibrary {positioning,shapes.misc}
\usetikzlibrary {decorations.shapes,shapes.geometric} 
\usetikzlibrary{graphs}

\newcommand{\ord}[1]{{\scriptscriptstyle (#1)}}

\title{\boldmath{Recursion for Differential Cross-Section 
\\
from the Optical Theorem}}

\author[a]{Vatsal Garg}
\author[b]{Hojin Lee}
\author[c,d]{Kanghoon Lee}

\affiliation[a]{Indian Institute of Science Education and Research, Pune, Dr. Homi Bhabha Road, Pashan, Pune - 411008, India}
\affiliation[b]{Department of Physics and Astronomy, Seoul National University, Seoul 08826, Korea}
\affiliation[c]{Asia Pacific Center for Theoretical Physics, Postech, Pohang 37673, Korea}
\affiliation[d]{Department of Physics, Postech, Pohang 37673, Korea}

\emailAdd{vatsal.garg@students.iiserpune.ac.in}
\emailAdd{zet4gra9er@snu.ac.kr}
\emailAdd{kanghoon.lee1@gmail.com}

\abstract{We present a novel framework for computing differential cross-sections in quantum field theory using the optical theorem and loop amplitudes, circumventing the traditional method of squaring scattering amplitudes. This approach addresses two major computational challenges in high-multiplicity processes: complexity from amplitude squaring and the extensive summations over color and helicity. Our method employs quantum off-shell recursion, a loop-level generalization of Berends--Giele recursion, combined with Veltman's largest time equation (LTE) through a doubling prescription of fields. By deriving Dyson--Schwinger equations within this doubled framework and constructing quantum perturbiner expansions, we develop recursive relations for generating LTEs. We validate our method by successfully reproducing the differential cross-section for tree-level $2 \to 2$ and $2 \to 4$ scalar scattering for $\phi^{4}$ theory through one-loop and three-loop amplitude calculation respectively. This framework offers an efficient alternative to conventional methods and can be broadly applied to theories with color charges, such as QCD and the Standard Model.

}

\begin{document}
\bibliographystyle{JHEP}

\preprint{xxx}
\maketitle

\section{Introduction}

The cross-section is a fundamental physical observable in particle physics, providing a connection between theoretical predictions and experimental measurements. It quantifies the probability of specific scattering processes occurring during particle interactions, including decay rates, branching ratios, and the parton distribution functions. Within the framework of quantum field theory (QFT), the principle of unitarity and analyticity of S-matrix plays a fundamental role. The optical theorem \cite{Feenberg:1932zz,BohrPeierlsPlaczek} (See also \cite{NewtonOpticalTheorem} and references therein.), a direct implication of S-matrix unitarity, establishes an important connection between the imaginary part of the scattering amplitude and the cross-section, making it a powerful tool for both theoretical calculations and experimental predictions.

Although the differential cross-section is formally obtained by squaring the scattering amplitude, this procedure becomes particularly challenging in high-multiplicity processes involving color charges, even when the amplitudes are explicitly known. In general, the computation of such amplitudes experiences factorial growth of the number of terms as the number of external particles increases, and the squaring procedure leads to $ \mathcal{O}(N^2) $ complexity. In addition, it involves complicated sums over color and helicity states. These challenges make the calculations computationally impractical. While significant developments have been made in efficiently computing scattering amplitudes using techniques, the progress of computational methods for differential cross-sections has been relatively limited.

In this work, we introduce a new framework for calculating differential cross-sections based on the optical theorem, which allows us to derive the cross-section using loop amplitudes instead. For efficient computation of the loop amplitudes, we employ the quantum off-shell recursion \cite{Lee:2022aiu,Cho:2023kux}, which is a loop-level generalization of Berends and Giele's off-shell recursions for tree-level amplitude \cite{Berends:1987me}. The quantum off-shell recursion formalism is based on the perturbiner method \cite{Rosly:1996vr,Rosly:1997ap,Selivanov:1997aq,Selivanov:1997an,Selivanov:1997ts,Lee:2022aiu} and the Dyson--Schwinger equation, providing an efficient tool for constructing loop amplitudes by iteratively solving off-shell recursion relations.

At the perturbative level, the unitarity of the S-matrix can be systematically imposed on individual diagrams through the Cutkosky rules \cite{Cutkosky:1960sp}. While the Cutkosky rules provide a powerful method by replacing the imaginary part of a loop diagram with a set of cut diagrams, their application requires a detailed understanding of the analytic properties and cut structures of loop integrands, which become highly intricate as diagrams get complicated. Instead, we exploit Veltman's largest time equation (LTE) \cite{Veltman:1963th,tHooft:1973wag,Veltman:1994wz,HoratiuQFT}, which provides systematic rules for identifying the cuts by doubling the interaction vertices and Schwinger--Keldysh propagators \cite{Schwinger:1960qe,Keldysh:1964ud}. One of the key aspects of the LTE is that it can provide algebraic criteria for identifying the unitary cuts without considering explicit diagrams. This feature is crucial for our purpose because the loop amplitudes are derived by solving the algebraic recursion relations.

We begin by defining an action for LTE by doubling the field content for $\phi^4$ theory, $\phi^{+}$ and $\phi^{-}$ and incorporating Schwinger--Keldysh propagators. This prescription, we refer to the \emph{doubling prescription}, is the same setup as the effective action formalism for Schwinger--Keldysh formalism (See \cite{CHOU19851} and references therein). By virtue of the doubled fields, the doubled interaction vertices and the Schwinger--Keldysh propagators are identical to the vertices and propagators in LTE. We derive the Dyson--Schwinger equation in the context of the doubling prescription, which is a quantum generalization of the classical EoM.

Next, we construct the quantum off-shell recursions for generating the LTEs of the $\phi^4$ theory. To obtain the off-shell recursions within the doubling prescription, we use the Dyson--Schwinger (DS) equations corresponding to the doubled action and the perturbiner expansion for each doubled field variable. First, we derive the DS equation, a quantum counterpart of the classical EoM, by deforming the EoMs for the doubled field variables. We explicitly derive the DS equation up to the 3-loop order. Second, we construct the quantum perturbiner expansion from the one-point function in the presence of the pair of external sources within the quantum effective action formalism. By substituting the perturbiner expansion into the Dyson--Schwinger equation, we can derive the off-shell recursions for LTE. Here the three-loop amplitude corresponds to the $2\to4$ cross-section at the tree level.

Finally we solve the recursions and obtain amplitudes order by order in loop orders. We call the amplitude from the off-shell recursions as \emph{doubled amplitude}. Our results show that the solutions of the recursions in one-loop and three-loop orders precisely match the differential cross-section for the $2 \to 2$ and $2 \to 4$ scalar scattering process at the tree level. However, one can check that the doubled amplitude includes not only the target cross-section but also irrelevant terms corresponding to other cross-sections. To address this issue, we develop an algorithm to extract specific loop orders precisely within the doubled amplitude. 

Our method bypasses the squaring of amplitudes and the cumbersome sums over color and helicity, allowing for the direct computation of the differential cross-section. Moreover, we exploit the quantum off-shell recursion, which significantly reduces the computational workload. Thus, it offers an efficient alternative compared to conventional methods. Moreover, since this method can be applied to any theory with a given action, it can be applied extensively, particularly useful for theories with color charges, such as QCD or the Standard Model. 

The organization of this paper is as follows. In section \ref{Sec:2}, we introduce the doubling prescription with action and derive the DS equation. In section \ref{Sec:3}, we derive the perturbiner expansion and the QOR up to three-loop order. We also discuss how to extract a certain loop order selectively. In section \ref{Sec:4}, we solve the recursion relation and show that our method exactly reproduces the known results. Finally in section \ref{Sec:5}, we conclude with several future directions.


\section{Action formalism for the Largest Time Equation}\label{Sec:2}

The largest time equation (LTE) provides a systematic framework for describing the optical theorem by extending the Feynman rules. Within this approach, the imaginary part of a given diagram is represented through cut diagrams constructed using the enlarged propagators. Consequently, LTE offers an algebraic representation of the optical theorem.

In this section, we present an action formalism for LTE\footnote{See appendix \ref{App:A} for a concise review of the LTE.} by introducing a doubled field content. This formalism is identical to the effective action formalism for the Schwinger--Keldysh formalism \cite{CHOU19851}. Notably, the doubled action reproduces the extended Feynman rules precisely, providing a consistent theoretical tool for LTE. Furthermore, we derive the Dyson--Schwinger equation for the doubled action, which plays a key role in constructing the recursion relation for LTE in the following sections.

\subsection{The doubling prescription}

Let us consider the S-matrix, denoted as $\mathcal{S}$, for the $\phi^{4}$ theory. According to the LSZ reduction formula, we may represent the S-matrix elements in terms of the correlation functions. The correlation functions associated with $\mathcal{S}$ and $\mathcal{S}^{\dagger}$ are generated by $Z[j^{+}]$ and $Z^*[j^{-}]$ respectively. Here $j^{+}$ and $j^{-}$ are external sources related by the complex conjugation, $(j^{+})^{*} = j^{-}$. Similarly, the diagrams associated with $\mathcal{S} \mathcal{S}^{\dag}$, which turned out to be the identity element by the unitarity, are generated by the following generating functional $Z[j^{+},j^{-}]$
\begin{equation}
\begin{aligned}
  Z[j^{+},j^{-}]
  =
  Z[j^{+}] Z^{\ast}[j^{-}]
  =
  \int \mathcal{D}\phi^{+} \mathcal{D}\phi^{-} e^{\frac{i}{\hbar} S[\phi^{+},j^{+}]-\frac{i}{\hbar} S^{*}[\phi^{-},J^{-}]}\,,
\end{aligned}\label{gen_functional}
\end{equation}
where the $\phi^{4}$ theory action $S[\phi^{\pm},j^{\pm}]$ is given by
\begin{equation}
  S[\phi^{\pm},j^{\pm}] 
  =
  \int \mathrm{d}^{4}x \bigg[ 
  - \frac{1}{2} \phi^{\pm} (-\Box + m^{2}\pm i \epsilon) \phi^{\pm} 
  - \frac{\lambda}{4!} \big(\phi^{\pm}\big)^{4}
  \pm j^{\pm} \phi^{\pm}
  \bigg]
\label{}\end{equation}
and $S^{*}$ implies that the opposite $i\epsilon$ prescription is applied, $i\epsilon \to -i\epsilon$. By definition, $Z[j^{+},j^{-}]$ is the norm of the vacuum-to-vacuum transition amplitude in the presence of the source 
\begin{equation}
  Z[j^{+},j^{-}] = \lVert_{j_{+}}\langle 0_{+}| 0_{-} \rangle_{j_{-}}\rVert^{2} > 0\,,
\label{Z_norm}\end{equation}
where $|0^{+}\rangle_{j_{+}}$ and $|0^{-}\rangle_{j_{-}}$ are the out and in vacua. 

The generating functional for connected diagrams that contributes to the transfer matrix $\mathcal{T}$, $\mathcal{S} = 1 + i \mathcal{T}$, are given by
\begin{equation}
  e^{\frac{i}{\hbar}  W[j^{+},j^{-}]}=Z[j^{+},j^{-}]\,,
\label{}\end{equation}
Then the condition in \eqref{Z_norm} implies that $W[j^{+},j^{-}]$ is purely imaginary
\begin{equation}
  W[j^{+},j^{-}] = \mbox{Im} W[j^{+},j^{-}]\,.
\label{}\end{equation}
Thus the correlation functions generated by $W[j^{+},j^{-}]$ contain the imaginary part only. 

We introduce the doublet of the scalar fields $\phi^{A}_{x} = \phi^{A}(x)$ and their external sources $j^{A}_{x} = j^{A}(x)$ 
\begin{equation}
  \phi^{A}_x=\left\{\phi^{+}_x, \phi^{-}_x\right\}, 
  \qquad
  j^{A}_x=\left\{j^{+}_x, j^{-}_x\right\} \,.
\label{}\end{equation}
Then the total action\footnote{Hereafter we denote the integrations as 
\begin{equation*}
    \int_{x,y,\ldots} = \int \mathrm{d}^{4} x\,\mathrm{d}^{4}y\cdots\,,
    \quad
    \int_{p,q,\ldots} = \int\frac{\mathrm{d}^{4} p}{(2\pi)^4}\frac{\mathrm{d}^{4}q}{(2\pi)^4}\cdots\,.
\end{equation*}
}, which we refer to this as the \emph{doubled action}, is rewritten in terms of the doublets
\begin{equation}
  S[\phi^{A},j^{A}] = -\frac12 \int_{x,y} 
  	  \phi^{A}_x K^{A B}_{xy} \phi^{B}_y
  	- \frac{1}{4!} \int_{x}V^{A B C D} \phi^{A}_x \phi^{B}_x \phi^{C}_x \phi^{D}_x
  	+ \eta^{A B} j^{A}_x \phi^{B}_x
  \,.
\label{total_action}\end{equation}
Here $K^{AB}_{xy}$ is the kinetic operator
\begin{equation}
  K^{AB}_{xy}  = K^{AB}(x-y)
  = 
  \begin{pmatrix} 
  \big(-\Box + m^{2} - i\epsilon\big)\delta^4(x-y) & 0 
  \\
  0 & - \big(-\Box + m^{2} + i\epsilon\big) \delta^4(x-y)
  \end{pmatrix}\,,
\label{}\end{equation}
and $V^{ABCD}$ are the interaction vertices
\begin{equation}
    V^{ABCD} = \begin{cases}
  	V^{++++} = \lambda \,,
  	\\
  	V^{----} = -\lambda \,, 
  	\\
  	0 \quad\text{for others}\,.
  \end{cases}
\label{}\end{equation}
Here $\lambda$ is the coupling constant. Finally, $\eta^{AB}$ is defined by
\begin{equation}
  \eta^{AB}
  = \begin{pmatrix} \eta^{++} & \eta^{+-} \\ \eta^{-+} & \eta^{--} \end{pmatrix}
  = \begin{pmatrix} 1 & 0 \\ 0 & -1\end{pmatrix} \,.
\end{equation}

In the doubled action \eqref{total_action}, the $+$ and $-$ sectors are completely decoupled if we use the diagonal propagator without mixing the $+$ and $-$ components 
\begin{equation}
  D^{AB}_{{\rm diag},xy} 
  = 
  \begin{pmatrix}
  D^{++}_{xy} & 0 
  \\ 
  0 & D^{--}_{xy}
  \end{pmatrix}\,.
\label{diag_propagator}\end{equation}
In this case any correlation functions with mixed $\phi_{+}$ and $\phi_{-}$ fields always vanish. However, we may choose another propagator using different boundary conditions, such as $\phi_{+}(t\to\infty) = \phi_{-}(t\to\infty)$ in Schwinger--Keldysh formalism, we may have nontrivial correlators. This exactly corresponds to the action formalism for the Schwinger--Keldysh formalism \cite{CHOU19851}. 

Let us define the propagator $D^{AB}_{xy}$ as the inverse of the kinetic operator $K^{AB}_{xy}$,
\begin{equation}
\begin{aligned}
  D^{AB}_{xy} 
  = 
  \begin{pmatrix}
  D^{++}_{xy} & D^{+-}_{xy} 
  \\ 
  D^{-+}_{xy} & D^{--}_{xy}
  \end{pmatrix}
  = 
  \begin{pmatrix}
  D_{F}(x-y) & - \Delta^{+}(x-y)
  \\ \Delta^{-}(x-y) & - D^{*}_{F}(x-y)
  \end{pmatrix}
\end{aligned}\,,\label{Propagator_matrix}
\end{equation}
where $D_{F}(x-y)$ and $D^{*}_{F}(x-y)$ are the conventional Feynman propagator and its complex conjugation, and $\Delta^{\pm}(x-y)$ are Wightman functions
\begin{equation}
  \Delta^{ \pm}(x-y)
  = \int_k \, e^{\pm ik \cdot (x-y)} \hat{\delta}(k^2+m^2) \, \Theta(k^0) 
  = \int \frac{\mathrm{d}^{3}\vec{k}}{(2 \pi)^{3} 2 E_{k}} e^{ \pm i k \cdot (x-y)}\,,
\label{Adv_Rev_Propagators}\end{equation}
where $\hat{\delta}(x) = 2\pi \delta(x)$ and $\Theta\left(x^0 - y^0\right)$ is the Heaviside step function. These are not independent but related to themselves 
\begin{align}
  D_{F}(x-y) &=
  \Theta\left(x^{0}-y^{0}\right) \Delta^{+}(x-y)+\Theta\left(y^{0}-x^{0}\right) \Delta^{-}(x-y)\,,\label{FeynmanExplicit}
  \\
  D_{F}^{*}(x-y) &=
  \Theta\left(y^{0}-x^{0}\right) \Delta^{+}(x-y)+\Theta\left(x^{0}-y^{0}\right) \Delta^{-}(x-y)\,,\label{AntiFeynmanExplicit}
\end{align}
where 
\begin{equation}
  \left(\Delta^{+}(x-y)\right)^{*}=\Delta^{+}(y-x) = \Delta^{-}(x-y)\,.
\label{}\end{equation}
It is straightforward to show that $D^{AB}_{xy}$ satisfies the inverse relation
\begin{equation}
    \int_y D^{A B}_{xy} K^{BC}_{yz} = \delta^{AC}\delta^4(x-z)\,.
\label{InversionRelation}\end{equation}
Note that this equality holds even in the presence of the off-diagonal part of $D^{AB}$ due to the on-shell condition in $\Delta^{{\pm}}$ defined in \eqref{Adv_Rev_Propagators}, $(k^{2}+m^{2})\hat{\delta}(k^{2}+m^{2}) = 0$. We refer to the doubled action together with the propagators in Schwinger--Keldysh formalism as the \emph{doubling prescription}.

\subsection{Largest Time Equation and Cuts}

The propagator choice in \eqref{Propagator_matrix} leads to a nontrivial deformation to the original theory. Here, the original theory means the imaginary part of the undoubled theory or, equivalently, the doubled action with the diagonal propagator \eqref{diag_propagator}. Let us denote the connected generating functional in the doubling prescription as $\mathcal{W}[j^{+},j^{-}]$. We also denote the $n$-point connected correlation functions for the original theory as $G_{c}(x_{1},x_{2},\cdots,x_{n})$ and for the deformed theory as
\begin{equation}
  G_{c}^{A_{1},A_{2},\cdots,A_{n}}(x_{1},x_{2},\cdots, x_{n})
  =
  \left\langle 0|T \phi^{A_{1}}(x_{1}) \phi^{A_{2}}(x_{2}) \cdots \phi^{A_{n}}(x_{n})|0\right\rangle_{c}\,.
\label{}\end{equation}

Since there are nontrivial interactions between $\phi^{+}$ and $\phi^{-}$ in the doubling prescription, $\mathcal{W}[j^{+},j^{-}]$ is decomposed as follows:
\begin{equation}
\begin{aligned}
  \mathcal{W}[j^{+},j^{-}] 
  &= 
  W[j^{+}] - W^{*}[j^{-}] + \tilde{W}[j^{+},j^{-}]\,,
  \\
  &= 
  W[j^{+},j^{-}] + \tilde{W}[j^{+},j^{-}]\,,
\end{aligned}\label{}
\end{equation}
where $\tilde{W}[j^{+},j^{-}]$ denotes the terms with mixed $j^{+}$ and $j^{-}$ and generates correlation functions of mixed $\phi^{+}$ and $\phi^{-}$ fields. The imaginary part of the original correlation function $\mbox{Im}\,G_{c}(x_{1},x_{2},\cdots,x_{n})$ can be represented by $G_{c}^{A_{1},A_{2},\cdots,A_{n}}(x_{1},x_{2},\cdots, x_{n})$. According to the LTE\footnote{For a concise review, see appendix \ref{App:A}}, $\mathcal{W}[j^{+},j^{-}]-\tilde{W}[j^{+},j^{-}]$ should generate the righthand side of \eqref{LTE}, which is known as cut diagrams. Thus we compute $\mbox{Im}\,G_{c}(x_{1},x_{2},\cdots,x_{n})$

Although we do not use any diagram in actual computation for differential cross-sections, it is convenient to introduce Feynman rules for the doubled action to compare with the conventional language. The interaction vertices $V^{++++}$ and $V^{----}$ are denoted by the uncircled and circled diagrams
\begin{center}
\begin{tikzpicture}[scale=0.8]
  	\draw (0,0) -- (2,2);
  	\filldraw[black] (1,1) circle (2pt) node[anchor=west]{\qquad~~: $\displaystyle -i \lambda \int_x$} node[below=6pt]{$x$};
  	\draw (2,0) -- (0,2);
\end{tikzpicture}
\qquad\qquad
\begin{tikzpicture}[scale=0.8]
  	\draw (0,0) -- (2,2);
  	\filldraw[black] (1,1) circle (2pt) node[anchor=west]{\qquad~~: $\displaystyle i \lambda \int_x$};
  	\draw (1,1) circle [radius=5pt]  node[below=6pt]{$x$};
  	\draw (2,0) -- (0,2);
\end{tikzpicture}
\end{center}
Thus, the uncircled vertex is identical to the one in the conventional $\phi^{4}$ theory, and the circled vertex is its complex conjugate. 

Furthermore, the previous propagators, $D_{F}$, $\Delta^{\pm}$ and $D^{*}_{F}$, can be understood as the two-point functions connecting the two kinds of vertices
\begin{center}
\begin{tikzpicture}[scale=0.8]
  	\draw (0,0) -- (2,0);
  	\filldraw[black] (0,0) circle (2pt) node[below=6pt]{$x$};
  	\filldraw[black] (2,0) circle (2pt) node[below=6pt]{$y$} node[anchor=west]{~~~ $D_{F}(x-y)$\,,};
\end{tikzpicture}
\qquad\qquad
\begin{tikzpicture}[scale=0.8]
  	\draw (0,0) -- (2,0);
  	\filldraw[black] (0,0) circle (2pt) node[below=6pt]{$x$};
  	\filldraw[black] (2,0) circle (2pt) node[below=6pt]{$y$} node[anchor=west]{~~~ $\Delta^{+}(x-y)$\,,};
  	\draw (0,0) circle [radius=5pt];
        \filldraw[black] (1.2,0) -- (0.9,-0.17) -- (0.9,0.17) -- (1.2,0);
\end{tikzpicture}
\\
\begin{tikzpicture}[scale=0.8]
  	\draw (0,0) -- (2,0);
  	\filldraw[black] (0,0) circle (2pt) node[below=6pt]{$x$};
  	\filldraw[black] (2,0) circle (2pt) node[below=6pt]{$y$} node[anchor=west]{~~~ $\Delta^{-}(x-y)$\,,};
  	\draw (2,0) circle [radius=5pt];
        \filldraw[black] (0.8,0) -- (1.1,-0.17) -- (1.1,0.17) -- (0.8,0);
\end{tikzpicture}
\qquad\qquad
\begin{tikzpicture}[scale=0.8]
  	\draw (0,0) -- (2,0);
  	\filldraw[black] (0,0) circle (2pt) node[below=6pt]{$x$};
  	\filldraw[black] (2,0) circle (2pt) node[below=6pt]{$y$} node[anchor=west]{~~~ $D^{*}_{F}(x-y)$\,.};
  	\draw (2,0) circle [radius=5pt];
 	\draw (0,0) circle [radius=5pt];
\end{tikzpicture}
\end{center}
where the arrows denote energy flows, which is clear in the momentum space. 

Consider the momentum space propagator $D^{AB}_{p}$
\begin{equation}
  \tilde{D}^{AB}_{p} 
  =
  \begin{pmatrix}
  \tilde{D}^{+} (p) & - \tilde{\Delta}^{+} (p)
  \\
  \tilde{\Delta}^{-} (p) & - \tilde{D}^{-} (p)
  \end{pmatrix}
  =
  \begin{pmatrix} 
  \dfrac{1}{p^{2}+m^{2} - i \varepsilon} &  - \hat{\delta}(p^{2}+m^{2}) \Theta(p^{0}) 
  \\ 
  \hat{\delta}(p^{2}+m^{2}) \Theta(-p^{0}) & -\dfrac{1}{p^{2}+m^{2} + i \varepsilon}
  \end{pmatrix}\,,
\label{}\end{equation}
and $\tilde{\Delta}^{\pm}(p)$ automatically define the cuts due to the delta-function $\hat{\delta}(p^{2}+m^{2})$, by which intermediate particles go on-shell,
\begin{equation}
\begin{aligned}
  \begin{tikzpicture}[scale=0.8]
  	\draw (0,0) -- (2,0);
  	\filldraw[black] (0,0) circle (2pt) node[below=6pt]{$x$} node[anchor=east]{$\Delta^{+}(x-y)$~:~~~~~};
  	\filldraw[black] (2,0) circle (2pt) node[below=6pt]{$y$} node[anchor=west]{~~~$\Longrightarrow$};
  	\draw (0,0) circle [radius=5pt];
  	\draw (4,0) -- (6,0);
  	\filldraw[black] (4,0) circle (2pt) node[below=6pt]{$x$};
  	\filldraw[black] (6,0) circle (2pt) node[below=6pt]{$y$};
  	\draw (4,0) circle [radius=5pt];
  	\draw[thick,dashed,red] (5,-0.4) -- (5, 0.4);
\end{tikzpicture}
\\
\begin{tikzpicture}[scale=0.8]
  	\draw (0,0) -- (2,0);
  	\filldraw[black] (0,0) circle (2pt) node[below=6pt]{$x$} node[anchor=east]{$\Delta^{-}(x-y)$~:~~~~~};
  	\filldraw[black] (2,0) circle (2pt) node[below=6pt]{$y$} node[anchor=west]{~~~$\Longrightarrow$};
  	\draw (2,0) circle [radius=5pt];
  	\draw (4,0) -- (6,0);
  	\filldraw[black] (4,0) circle (2pt) node[below=6pt]{$x$};
  	\filldraw[black] (6,0) circle (2pt) node[below=6pt]{$y$};
  	\draw (6,0) circle [radius=5pt];
  	\draw[thick,dashed,red] (5,-0.4) -- (5, 0.4);
\end{tikzpicture}
\end{aligned}\label{Cut_propagators}
\end{equation}
Here the red lines represent the cuts. We algebraically define the term \emph{cutting} as substituting the value of $\tilde{\Delta}^{\pm}$
\begin{equation}
  \text{cutting}: \quad \tilde{\Delta}^{\pm}(p) \to \hat{\delta}(p^{2}+m^{2}) \Theta(\pm p^{0})\,.
\label{Def_cutting}\end{equation}
%

\subsection{Dyson--Schwinger equation in doubling prescription}

Dyson--Schwinger (DS) equation is regarded as the quantum counterpart of the classical EoM. It is an infinitely coupled PDE and contains nonperturbative information about QFT. It plays a crucial role in deriving the quantum off-shell recursions in the next section. A priori DS equation provides relations between correlation functions, but we may rewrite it in terms of the classical field $\varphi^A_{x}$, a one-point function in the presence of the external source $j^{A}_{x}$
\begin{equation}
  \varphi^{A}_{x} 
  = 
  \eta^{AB} \frac{\delta W[j_{+},j_{-}]}{\delta j^{B}_{x}} \,,
\label{classical_field}\end{equation}
where $W[j_{+},j_{-}]$ is the generating functional for connected diagrams. We also define \emph{descendant fields} \cite{Lee:2022aiu,Cho:2023kux}
\begin{equation}
\begin{aligned}
  \psi^{AB}_{xy} 
  &= \frac{\delta \varphi^A_x}{\delta j^{B}_y}\,, 
  \qquad
  \psi^{ABC}_{xyz} 
  &= \frac{\delta^{2}\varphi^{A}_{x}}{\delta j^{B}_y\delta j^{C}_z}\,,
  \qquad
  \psi^{ABCD}_{xyzw} 
  &= \frac{\delta^{3}\varphi^{A}_{x}}{\delta j^{B}_y\delta j^{C}_z\delta j^{D}_w} \,,
  \quad
  \text{etc.}
\end{aligned}\label{DescendantFields}
\end{equation}

We can derive the DS equation by promoting the fields $\phi^{A}$ in classical EoM to the following operator $\hat{\phi}^{A}$ \cite{Brown:1992db}
\begin{equation}
  \phi^{A}\to \hat{\phi}^{A} = \varphi^{A} + \frac{\hbar}{i}\frac{\delta}{\delta j^{A}}\,,
\label{deformation}\end{equation}
where $\varphi^{A}_{x}$ is the classical field in \eqref{classical_field}.

For the doubled action in \eqref{total_action}, the classical EoM is given by
\begin{equation}
  -\int_y K^{AB}_{xy}\phi^B_y - \frac{1}{3!}V^{ABCD}\phi^B_x\phi^C_x\phi^D_x + \eta^{AB}j^B_x = 0\,.
\label{Classical_Equation_of_Motion}\end{equation}
Substituting the deformation \eqref{deformation} into the above EoM, we obtain DS equation in the doubling prescription
\begin{equation}
  \varphi^{A}_{x}
  =
  \int_{y}
    D^{AE}_{xy} \bigg[
  - \frac{1}{3!}V^{BCDE} \Big(\varphi^B_y\varphi^C_{y}\varphi^D_{y}
  + 3\frac{\hbar}{i}\varphi^{(B}_{y}\psi^{CD)}_{yy} - \hbar^2\psi^{(BCD)}_{yyy}\Big)
  + \eta^{BE}j^B_{y}
  \bigg] \,.
\label{}\end{equation}
It is convenient to use the components explicitly
\begin{equation}
\begin{aligned}
  \varphi^A_x
  &=
  \int_y D^{A+}_{xy} \bigg[
  	  j^+_y 
  	- \frac{\lambda}{3!}\big(\varphi^+_y\big)^3 
  	- \frac{\lambda}{2!}\frac{\hbar}{i}\varphi^+_y\psi^{++}_{yy}
  	+ \frac{\lambda}{3!}\hbar^2\psi^{+++}_{yyy} \bigg]
  \\&\quad
  - \int_y D^{A-}_{xy} \bigg[
  	  j^-_y 
  	- \frac{\lambda}{3!}\big(\varphi^-_y\big)^3
  	- \frac{\lambda}{2!}\frac{\hbar}{i}\varphi^-_y\psi^{--}_{yy}
  	+ \frac{\lambda}{3!}\hbar^2\psi^{---}_{yyy}\bigg] \,.
\end{aligned}\label{ExplicitDysonSchwinger}
\end{equation}
We should also derive the DS equation for the descendant fields, which is derived by taking the functional derivatives on \eqref{ExplicitDysonSchwinger} with respect to the sources $j^A_x$. 

In general, this procedure does not stop in finite order and generates an infinite number of descendant fields and their equations. One way to terminate this is to introduce the $\hbar$ expansion. Once we designate a certain order in $\hbar$, then we can truncate the higher descendant fields because they always carry higher $\hbar$ orders in DS equation. We substitute the following $\hbar$ expansion into the DS equation and its descendants
\begin{equation}
\begin{aligned}
  \varphi^{A}_x &= \sum_{n=0}^{\infty} \left(\frac{\hbar}{i}\right)^{n} \varphi^{A}_{\ord{n},x}\,, 
  \qquad
  \psi^{AB}_{xy} = \sum_{n=0}^{\infty} \left(\frac{\hbar}{i}\right)^{n} \psi^{AB}_{\ord{n},xy}\,, 
  \\
  \psi^{ABC}_{xyz} &= \sum_{n=0}^{\infty} \left(\frac{\hbar}{i}\right)^{n} \psi^{ABC}_{\ord{n},xyz}\,,
  \qquad
  \text{etc}\,.
\end{aligned}\label{HbarExpansions_fields}
\end{equation}
We will truncate the DS equation up to $\hbar^{3}$ order or the three-loop order. One can find the explicit form of the DS equations for descendant fields in appendix \ref{App:B}.

\section{Off-shell recursions for differential cross-sections}\label{Sec:3}

In general, the Quantum Off-shell Recursion (QOR) can be systematically constructed from the perturbiner method using the Dyson--Schwinger equation. In this section, we derive the quantum perturbiner expansion for the doubling prescription, which is a generating function for the off-shell currents for the imaginary part of the amplitude. We then derive the QOR for the LTE by substituting the perturbiner expansion into the Dyson--Schwinger (DS) equations derived in the previous section. Finally, we discuss how to extract only the desired loop order accurately.

\subsection{Off-shell currents and the Largest Time Equation}

The off-shell current is defined by a collection of all possible Feynman diagrams associated with fixed external legs \cite{Berends:1987me}. Among the external legs, one of them is the off-shell leg, while all the others are amputated on-shell legs.  Recently the off-shell current has been extended to include diagrams at any loop orders \cite{Lee:2022aiu}\footnote{See also \cite{Gomez:2022dzk,Chen:2023bji,Cohen:2022uuw,Cohen:2023ekv,Gomez:2024xec}.}. We want to identify the off-shell currents with the amplitudes in the doubling prescription, which we shall refer to as \emph{doubled amplitudes}. We focus on the $2\to 2$ doubled amplitude at $(n-1)$-loop order, $\mathcal{A}^{{\scriptscriptstyle(n-1)}}_{1^{+}2^{-}3^{-}4^{+}}$, which is related to the tree level $2\to n$ differential cross-section of the original theory according to the optical theorem
\begin{equation}
  \mathcal{A}_{1^{+}2^{-}3^{-}4^{+}} = \sum_{X} \mathcal{M}(1^{+}4^{+}\to X) \Big(\mathcal{M}(2^{-}3^{-}\to X)\Big)^{*}
\label{}\end{equation}
where $X$ denotes the all possible external particles in $2\to n$ scattering. If we glue the external lines for $1^{+}$ and $2^{-}$ by identifying there external momenta $k_{1}=k_{2}$, as well as $3^{-}$ and $4^{+}$ case, we have the differential cross-section $\mathrm{d} \sigma_{2\to n}$ for the original $\phi^{4}$ theory.
\begin{figure}[t]
\begin{center}
\begin{tikzpicture}[scale=0.7]
  \node[] at (0,3) [left] {$1^{+}$};
  \node[] at (0,0) [left] {$4^{+}$};
  \node[] at (3,3) [right] {$2^{-}$};
  \node[] at (3,0) [right] {$3^{-}$};
  \node[] at (-1.5,1.5) [left] {\large $\mathcal{A}^{{\scriptscriptstyle(n-1)}}_{1^{+}2^{-}3^{-}4^{+}}~=$};
  \draw[thick] (0,0) -- (3,3);
  \draw[thick] (3,0) -- (0,3);
  \filldraw[fill=lightgray!,thick] (1.5,1.5) circle (30pt);
  \node[] at (4,1.5) [right] {\Large$=$};

  \node[] at (6,3) [left] {$1^{+}$};
  \node[] at (6,0) [left] {$4^{+}$};
  \draw[thick] (6,0) -- (7.5,1.5);
  \draw[thick] (6,3) -- (7.5,1.5);
  \draw[thick] (7.5,1.5) to[out=60, in =180] (9.5,3);
  \draw[thick] (7.5,1.5) to[out=40, in =180] (9.5,2.5);
  \draw[thick] (7.5,1.5) to[out=20, in =180] (9.5,2);
  \draw[thick] (7.5,1.5) to[out=-20, in =180] (9.5,1);
  \draw[thick] (7.5,1.5) to[out=-40, in =180] (9.5,0.5);
  \draw[thick] (7.5,1.5) to[out=-60, in =180] (9.5,0);
  \filldraw[fill=lightgray!,thick] (7.5,1.5) circle (30pt);
  \node[] at (7.5,1.5) [] {$\mathcal{M}_{2\to n}$};
  \node[] at (9.5,1.5) [] {$\cdots$};
  \draw[thick,dashed,red] (9.5,-0.5) -- (9.5, 3.5);
  \draw[thick] (9.5,3) to[out= 0, in =120] (11.5,1.5);
  \draw[thick] (9.5,2.5) to[out=0, in =140] (11.5,1.5);
  \draw[thick] (9.5,2) to[out=0, in =160] (11.5,1.5);
  \draw[thick] (9.5,1) to[out=0, in =-160] (11.5,1.5);
  \draw[thick] (9.5,0.5) to[out=0, in =-140] (11.5,1.5);
  \draw[thick] (9.5,0) to[out=0, in =-120] (11.5,1.5);
  \draw[thick] (11.5,1.5) -- (13,3);
  \draw[thick] (11.5,1.5) -- (13,0);
  \filldraw[fill=lightgray!,thick] (11.5,1.5) circle (30pt);  
  \node[] at (11.5,1.5) [] {$\mathcal{M}^{*}_{2\to n}$};
  \node[] at (13,3) [right] {$2^{-}$};
  \node[] at (13,0) [right] {$3^{-}$};
\end{tikzpicture}
\end{center}\caption{Optical theorem in the doubling prescription}\label{optical_theorem}
\end{figure}
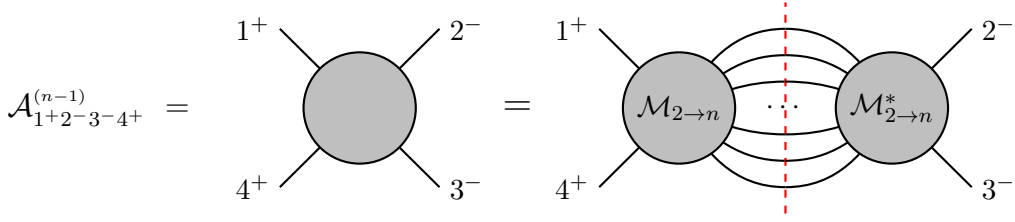

If we label the external legs as $i= 1,\cdots, N$, the off-shell current in doubling prescription $\Phi^A_{i_1\cdots i_N}$ is defined as
\begin{equation}
  \Phi^A_{i_1\cdots i_N} 
  =
  \tilde{G}^{A B_{1}B_{2}\cdots B_{N}}_c(-k_{12\cdots N},k_{1},k_{2},\cdots,k_{N}) \prod_{n=1}^N\left(\frac{i}{\hbar}\right)
  \eta^{B_{n} C_{n}}\tilde{\mathbf{K}}^{C_n D_n}\left(k_{i_{n}}\right)\epsilon^{D_{n}}_{i_{n}}\,,
\label{OffShellCurrent}\end{equation}
where $\tilde{G}^{A B_{1}B_{2}\cdots B_{n}}_c$ is the momentum space correlation function in the doubling prescription, and $k_{i}$ are the external momenta. $\tilde{\mathbf{K}}^{AB}(p)$ is the inverse of dressed propagator $\tilde{\mathbf{D}}^{AB}(p)$ for the amputation of the external legs. It satisfies the inverse relation $\tilde{\mathbf{K}}^{AB}(p)\tilde{\mathbf{D}}^{BC}(p) = \delta^{AC}$
\begin{equation}
  \tilde{\mathbf{D}}^{AB} (p) 
  = 
  \frac{i}{\hbar}\int_x e^{-ip \cdot (x-y)} \langle0\!\left|T\phi^A_x\phi^B_y\right|\! 0 \rangle\,,
\label{DressedPropagator}\end{equation}
and $\epsilon^C_i$ is a polarization determining whether the $i$-th leg corresponds to the field $\phi^+$ or $\phi^-$
\begin{equation}
    \epsilon^A_i = \begin{pmatrix}
        \epsilon^+_i \\
        \epsilon^-_i
    \end{pmatrix}\,.\label{Polarization_vec}
\end{equation}
For instance, by setting 
\begin{equation}
     \epsilon^A_1 = \begin{pmatrix}
        1 \\
        0
    \end{pmatrix}\,,
\end{equation}
we specify that the external field at the first leg is $\phi^+$. 

\begin{figure}[t]
\begin{center}
\begin{tikzpicture}[scale=0.7]
  \node[] at (0,0) [left] {$3^{-}$};
  \node[] at (1.5,-0.8) [right] {$2^{-}$};
  \node[] at (3,0) [right] {$1^{+}$};
  \node[] at (-1.5,1.5) [left] {\large $\Phi^{+}_{1^{+}2^{-}3^{-}}~=$};
  \draw[ultra thick] (1.5,1.5) -- (1.5,3.8);
  \draw[thick] (1.5,1.5) -- (3,0);
  \draw[thick] (1.5,1.5) -- (1.5,-0.8);
  \draw[thick] (1.5,1.5) -- (0,0);
  \filldraw[fill=lightgray!,thick] (1.5,1.5) circle (30pt);
\end{tikzpicture}
\end{center}
\caption{The off-shell current corresponding to $\mathcal{A}_{1^{+}2^{-}3^{-}4^{+}}$. The thick line is the off-shell leg.}
\label{OffShellCurrentFigure}\end{figure}
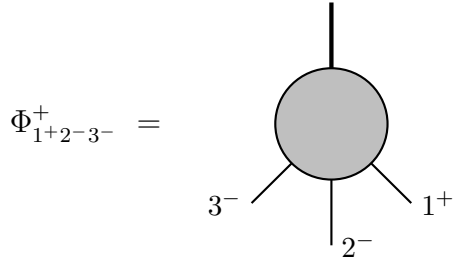

To derive the doubled amplitudes from the off-shell current, we need to amputate the off-shell leg and take the on-shell limit, $k_{1\cdots N}^2 + m_\mathrm{phys}^2 \to 0$, 
\begin{equation}\label{Amputation}
\mathcal{A}^{(n)}_{12\cdots (N+1)} = \lim_{k_{1\cdots N}^2 \to - m_\mathrm{phys}^2}\frac{i}{\hbar} \eta^{AB}\epsilon^{A}_{N+1}\hat{\Phi}^{{\scriptscriptstyle(n)},B}_{1\cdots N}\,,
\end{equation}
where $\hat{\Phi}^{{\scriptscriptstyle(n)},A}_{1\cdots N}$ is the $n$-loop amputated current
\begin{equation}
  \hat{\Phi}^{{\scriptscriptstyle(n)},A}_{1\cdots N} 
  = 
  \sum_{m=0}^n \tilde{\mathbf{K}}^{AB} \left(k_{1\cdots N}\right) \Phi^{B}_{1\cdots N}\Big|_{\mathcal{O}(\hbar^{n})}\,.
\label{}\end{equation}
Then the off-shell current corresponding doubled amplitude $\mathcal{A}^{{\scriptscriptstyle(n-1)}}_{1^{+}2^{-}3^{-}4^{+}}$ is given by $\Phi^{{\scriptscriptstyle(n-1)},+}_{1^{+}2^{-}3^{-}}$, which is depicted in figure \ref{OffShellCurrentFigure}.
%

\subsection{Perturbiner expansion for the Largest Time Equation}

The most important property of the off-shell current is that it satisfies a recursion relation. Berends and Giele constructed the recursions exploiting the hierarchical structure in tree diagrams \cite{Berends:1987me}, which is always applicable when there are a finite number of interaction vertices. Recently, the off-shell recursion relation, which previously existed only at the tree level, has been extended to the loop level, which is the so-called quantum off-shell recursion. It is obtained algebraically using the perturbiner method and DS equation without relying on the direct structure of the diagrams. 
 
 The key object in the perturbiner method is the perturbiner expansion, which serves as a generating function for off-shell currents. The expansion can be derived from the classical field, the one-point function with external sources, by substituting the appropriate form of $j^{A}_{x}$. The quantum off-shell recursion is derived by substituting the perturbiner expansion into the DS equation.

Let us now investigate what this quantum off-shell current corresponds to under the doubling prescription. We may expand the generating functional $W[j^{+},j^{-}]$ formally with respect to $j^{A}_{x}$
\begin{equation}
\begin{aligned}
  W[j^{A}] &=
  \left.\sum_{n=2}^{\infty} \frac{1}{n !} \int_{x_{1},x_{2}, \cdots, x_{n}} \frac{\delta^{n} W[j^{A}]}{\delta j^{A_{1}}_{x_{1}} \delta j^{A_{2}}_{x_{2}} \cdots \delta j^{A_{n}}_{x_{n}}}\right|_{j^{A}=0} 
  	j^{A_{1}}_{x_{1}} j^{A_{2}}_{x_{2}} \cdots j^{A_{n}}_{x_{n}}\,,
  \\&=
  -i \hbar \sum_{n=2}^{\infty} \frac{1}{n !} \int_{x_{1},x_{2}, \cdots,x_{n}} 
  G_{c}^{A_{1},A_{2},\cdots,A_{n}}(x_{1},x_{2},\cdots, x_{n})
  \frac{i j^{A_{1}}_{x_{1}}}{\hbar} \frac{i j^{A_{2}}_{x_{2}}}{\hbar} \cdots \frac{i j^{A_{n}}_{x_{n}}}{\hbar}\,,
\end{aligned}\label{FDGenerating_functional}
\end{equation}
and the classical field $\varphi^{A}_{x}$ in \eqref{classical_field} can also be expanded similarly,
\begin{equation}
  \varphi^{A}_{x}
  =
  \sum_{n=1}^{\infty} \frac{1}{n !} \int_{y_{1}, y_{2}, \cdots, y_{n}} 
  G_{c}^{A,A_{1},\cdots,A_{n}}(x,y_{1},\cdots, y_{n})_{c}
  \frac{i j^{A_{1}}_{y_{1}}}{\hbar} \frac{i j^{A_{2}}_{y_{2}}}{\hbar} \cdots \frac{i j^{A_{n}}_{y_{n}}}{\hbar}\,.
\label{ClassicalFieldExpansion}\end{equation}
Note that the external sources are completely arbitrary so far. To relate with the amplitudes, we identify $j^A_x$ as the inverse of the dressed two-point function for amputation, according to the Lehmann--Symanzik--Zimmermann (LSZ) reduction formula
\begin{equation}
  j^A_x = \sum_{i=1}^{N} \eta^{AB} \tilde{\mathbf{K}}^{BC} (-k_i) \epsilon^C_i e^{-i k_i \cdot x}\,.
\label{ExtSourceDefinition}\end{equation}

Substituting \eqref{ExtSourceDefinition} into the expansion of $\varphi^{A}_{x}$ in \eqref{ClassicalFieldExpansion}, we have the perturbiner expansion for the doubling prescription 
\begin{equation}
    \varphi^{A}_{x} = \sum_{\mathcal{P}}\Phi^{A}_{\mathcal{P}}e^{-ik_{\mathcal{P}} \cdot x}\,,
\label{CurrentPerturbiner}\end{equation}
where $\mathcal{P}$ is the word that labels the external multi-particle states and $k_\mathcal{P} = \sum_\ell k_\ell$ for $\ell \subseteq \mathcal{P}$, such as $k_{ijk} = k_{i}+k_{j}+k_{k}$.  We call the length of the words their rank and denote it as $|\mathcal{P}|$. Here $\sum_{\mathcal{P}}$ is a sum over all possible words. The single-particle states are labeled by letters, denoted by $i, j, k, l,\cdots$, whereas combinations of these states without ordering, or words, are represented as $\mathcal{P},\mathcal{Q},\mathcal{R}$, for instance, $\mathcal{P} = ijk$. 

Likewise, we can derive the perturbiner expansion of the descendant fields as follows:
\begin{equation}
\begin{aligned}
  \psi^{AB}_{x,y} &= \int_p \Psi^{AB}_{p|\emptyset}e^{ip \cdot (x-y)} + \sum_{\mathcal{P}}\int_p \Psi^{AB}_{p|\mathcal{P}} e^{ip \cdot (x-y)}e^{-ik_{\mathcal{P}} \cdot x}\,,
  \\
  \psi^{ABC}_{x,y,z} &= \sum_{\mathcal{P}}\int_{p,q} \Psi^{ABC}_{pq|\mathcal{P}} e^{i p \cdot (x-y) + iq \cdot (x-z)}e^{-ik_{\mathcal{P}} \cdot x}\,,
  \\
  \psi^{ABCD}_{x,y,z,w} &=  \int_{p,q,r}\Psi^{ABCD}_{pqr|\emptyset}e^{ip \cdot (x-y) + iq \cdot (x-z) + ir \cdot (x-w)}
  \\&\quad
  +\sum_{\mathcal{P}}\int_{p,q,r}\Psi^{ABCD}_{pqr|\mathcal{P}}e^{ip \cdot (x-y) + iq \cdot (x-z) + ir \cdot (x-w)}e^{-ik_{\mathcal{P}} \cdot x}\,,    
\end{aligned}\label{}
\end{equation}
where $\Psi^{AB}_{p|\mathcal{P}}, \Psi^{ABC}_{pq|\mathcal{P}}$ and $ \Psi^{ABCD}_{pqr|\mathcal{P}}$ are the quantum off-shell currents for descendant fields and hence are also referred to as the descendant currents.

As we have defined $\hbar$-expansion of the classical fields in \eqref{HbarExpansions_fields}, we derive the off-shell currents as
\begin{equation}
\begin{aligned}
  \Phi^{A}_{\mathcal{P}} &= \sum_{n=0}^{\infty} \int_{\ell_{1},\ell_{2},\cdots, \ell_{n}}\Phi^{\scriptscriptstyle{(n)}}_{\ell_{1}\cdots \ell_{n}} \big|_{\mathcal{P}}^{A}\bigg(\frac{\hbar}{i}\bigg)^{n}\,,
  \\
  \Psi^{A B_{1}\cdots B_{m}}_{\ell_{1}\ell_{2}\cdots \ell_{m}|\hat{\mathcal{P}}}
  &=
  \sum_{n=0}^{\infty} \int_{\ell'_{1},\ell'_{2},\cdots, \ell'_{n}} \Psi^{\scriptscriptstyle{(n)}}_{\ell'_{1} \cdots \ell'_{n}}\big|^{AB_{1}\cdots B_{m}}_{\ell_{1}\cdots \ell_{m}|\hat{\mathcal{P}}} \bigg(\frac{\hbar}{i}\bigg)^{n}\,,
\end{aligned}\label{current_loop_expansion}
\end{equation}
where $\ell_{1}, \ell_{2}, \cdots$ are the loop momenta. We also introduce an extended word $\hat{\mathcal{P}}$, where $\hat{\mathcal{P}} = \emptyset\cup\mathcal{P}$, because the descendant currents admit zero modes. Since the correlators are expanded in $\hbar$, it is obvious that the quantum off-shell currents are also expanded in $\hbar$.

In general, off-shell recursions involve many products of currents accompanied by sums over words. To simplify the equations and make the structure more transparent, we introduce a multilinear bracket \cite{Cho:2023kux}, which represents the summations over products of the currents
\begin{equation}
\begin{aligned}
  \big\lceil \mathcal{A}, \mathcal{B} \big\rfloor_{\mathcal{P}}^{\scriptscriptstyle{(n)}}
  &=
  \sum_{m=0}^{n}\sum_{\mathcal{P}=\mathcal{Q}\cup \mathcal{R}} \mathcal{A}^{\scriptscriptstyle{(n-m)}}_{\mathcal{Q}} \mathcal{B}^{\scriptscriptstyle{(m)}}_{\mathcal{R}}\,,
  \\
  \big\lceil \mathcal{A}, \mathcal{B}, \mathcal{C} \big\rfloor^{\scriptscriptstyle{(n)}}_{\mathcal{P}}
  &=
  \sum_{\substack{m,p=0\\ n\geq m+p}}^{n} \sum_{\mathcal{P}=\mathcal{Q}\cup \mathcal{R} \cup \mathcal{S}} \mathcal{A}^{\scriptscriptstyle{(n-m-p)}}_{\mathcal{Q}} \mathcal{B}^{\scriptscriptstyle{(m)}}_{\mathcal{R}} \mathcal{C}^{\scriptscriptstyle{(p)}}_{\mathcal{S}}\,,
\end{aligned}\label{}
\end{equation}
where the superscript labeling the $\hbar$-order is the total $\hbar$-order of the currents in the bracket, and $\sum_{\mathcal{P}=\mathcal{Q}\cup\mathcal{R}}$ and $\sum_{\mathcal{P}=\mathcal{Q}\cup\mathcal{R}\cup\mathcal{S}}$ are summations over all possible distributions of the letters of the ordered words $\mathcal{P}$ into non-empty ordered words $\mathcal{Q}$, $\mathcal{R}$ and $\mathcal{S}$. If any descendant currents are positioned in one of the entries of the bracket, the summation over the words is replaced with the hatted words, including the zero modes. By definition, the bracket is unordered, $\big\lceil \mathcal{A}, \mathcal{B} \big\rfloor_{\mathcal{P}} = \big\lceil \mathcal{B}, \mathcal{A} \big\rfloor_{\mathcal{P}}$.

Next, we define a differential operator $Q^{A}_{\mathcal{P}}$ acting on the bracket and call it the \textit{descend operator}. It maps an off-shell current and descendant currents as follows:
\begin{equation}
\begin{aligned}
  Q^{A}_{p} \Phi^{I}_{\mathcal{P}} &= \Psi^{IA}_{p|\mathcal{P}} \,,
  \\
  Q^{A}_{p} \Psi^{BC_{1}\cdots C_{n}}_{q_{1}\cdots q_{n}|\hat{\mathcal{P}}} &= \Psi^{BC_{1}\cdots C_{n}A}_{q_{1}\cdots q_{n}p|\hat{\mathcal{P}}}\,.
\end{aligned}\label{Delta_operator}
\end{equation}
We also denote the successive applications of the descend operator as
\begin{equation}
  Q^{A_{1}A_{2}\cdots A_{n}}_{\ell_{1}\ell_{2}\cdots\ell_{n} }\big[\bullet\big] 
  = 
  Q^{A_{n}}_{\ell_{n}}\bigg[\cdots Q^{A_{2}}_{\ell_{2}}\Big[Q^{A_{1}}_{\ell_{1}}\big[\bullet\big]\Big]\cdots\bigg]\,.
\label{}\end{equation}
We require that $Q^{A}_{p}$ satisfies the Leibniz rule inside the bracket,
\begin{equation}
\begin{aligned}
  Q^{A}_{p} \big\lceil \Phi^{B}, \Phi^{C}\big\rfloor_{\mathcal{P}} 
  &= 
    \big\lceil Q^{A}_{p} \Phi^{B}, \Phi^{C}\big\rfloor_{\mathcal{P}} 
  + \big\lceil \Phi^{B}, Q^{A}_{p} \Phi^{C}\big\rfloor_{\mathcal{P}}\,,
  \\&
  = \big\lceil  \Psi^{B,A}_{p}, \Phi^{C}\big\rfloor_{\mathcal{P}}
  + \big\lceil \Phi^{B}, \Psi^{CA}_{p}\big\rfloor_{\mathcal{P}}\,,
  \\
  &=
  \sum_{\mathcal{P}
  =
    \hat{\mathcal{Q}}\cup \hat{\mathcal{R}}} \Big( \Psi^{BA}_{p|\hat{\mathcal{Q}}} \Phi^{C}_{\hat{\mathcal{R}}} 
  + \Phi^{B}_{\hat{\mathcal{Q}}} \Psi^{CA}_{p|\hat{\mathcal{R}}} \Big)\,.
\end{aligned}\label{}
\end{equation}
Note that the words in the summation of the last line are replaced by the hatted words.

Finally, we adopt the following notation for propagators,
\begin{equation}
    \begin{aligned}
    \tilde{D}^{AB}_{\mathcal{P}} &= \tilde{D}^{AB}\left(-k_{\mathcal{P}}\right)\,,
    \\
	\tilde{D}^{AB}_{p|\mathcal{P}} &= \tilde{D}^{AB}\left(p -k_{\mathcal{P}}\right)\,,
	\\
	\tilde{D}^{AB}_{pq|\mathcal{P}} &= \tilde{D}^{AB}\left(p+q -k_{\mathcal{P}}\right)\,,
	\\
	\tilde{D}^{AB}_{p|\emptyset} &= \tilde{D}^{AB}\left(p\right)\,, \quad \textrm{etc.}
\end{aligned}\label{PropagatorNotation}
\end{equation}
where $\tilde{D}^{AB}(p)$ is the momentum space propagator. If $Q^{A}_{p}$ acts on $\tilde{D}^{AB}_{\mathcal{P}}$ and $\tilde{D}^{AB}_{p|\mathcal{P}}$, we have
\begin{equation}
  Q^{A}_{p} \tilde{D}^{BC}_{\mathcal{P}}
  =
  \tilde{D}^{BC}_{p|\mathcal{P}}Q^{A}_{p} \,,
  \qquad
  Q^{A}_{q} \tilde{D}^{BC}_{p|\mathcal{P}} 
  = 
  \tilde{D}^{BC}_{pq|\mathcal{P}}Q^{A}_{q} \,.
\label{}\end{equation}
%

\subsection{Constructing recursion relations and obtaining amplitudes}

The off-shell recursions can be derived by substituting the perturbiner expansions into the DS equations and their descendants. 

\subsubsection{Tree level}
We begin the construction with tree level off-shell recursion. Substituting the $\hbar$ expansion of the fields introduced in \eqref{HbarExpansions_fields} into the DS equations \eqref{ExplicitDysonSchwinger}, we collect the $\hbar^{0}$ order terms
\begin{equation}
  \varphi^{A}_{{\scriptscriptstyle(0)},x} 
  = 
    \int_{y} D^{A+}_{xy}\left(j^+_{y} - \frac{\lambda}{3!}\big(\varphi^+_{{\scriptscriptstyle(0)},y}\big)^3\right) 
  - \int_{y} D^{A-}_{xy}\left(j^-_{y} - \frac{\lambda}{3!}\big(\varphi^-_{{\scriptscriptstyle(0)},y}\big)^3\right)\,.
\label{TreeLevelDysonSchwinger}\end{equation}
Substituting the perturbiner expansions \eqref{CurrentPerturbiner} into the tree-level DS equations above, we have the tree-level recursion relations
\begin{equation}
  \Phi^{\scriptscriptstyle(0)}\big|^{A}_{\mathcal{P}} 
  =
  - \frac{\lambda}{3!}D^{A+}_{\mathcal{P}}\left\lceil\Phi^{+},\Phi^{+},\Phi^{+}\right\rfloor^{{\scriptscriptstyle(0)}}_{\mathcal{P}}
  + \frac{\lambda}{3!} D^{A-}_{\mathcal{P}}\left\lceil\Phi^{-},\Phi^{-},\Phi^{-}\right\rfloor^{{\scriptscriptstyle(0)}}_{\mathcal{P}}\,.
\label{RecursionPhi0}\end{equation}

We first identify the initial conditions of the recursion by specifying the rank-1 currents using the definition of the off-shell current  \eqref{OffShellCurrent},
\begin{equation}
\begin{aligned}
  \sum_{i=1}^N\Phi^{\scriptscriptstyle(0)}\big|^{A}_{i} e^{-ik_{i}\cdot x} 
  =
  \int_y \Big(D^{A+}_{xy}j^{+}_y - D^{A-}_{xy}j^{-}_y\Big)
  = \sum_{i=1}^N \epsilon^{A}_{i} e^{-ik_{i}\cdot x}\,.
\end{aligned}
\end{equation}
Comparing coefficients on both sides above gives the initial condition for tree level currents,
\begin{equation}
  \Phi^{\scriptscriptstyle(0)}\big|^{A}_{i} = \epsilon_{i}^A\,.
\label{}\end{equation}

\subsubsection{One-loop level}
Next, we construct one-loop level recursion relations. Substituting the $\hbar$ expansion \eqref{HbarExpansions_fields} into the DS equations \eqref{ExplicitDysonSchwinger} and collecting the $\hbar^{1}$ order terms, we have
\begin{equation}
\begin{aligned}
  \varphi^A_{{\scriptscriptstyle(1)},x}
  &= 
  \int_{y} \left[ 
  	  D^{A+}_{xy}\left(j^{+}_{{\scriptscriptstyle(1)},y} 
  	- \frac{\lambda}{2!} \left(\varphi^+_{{\scriptscriptstyle(0)},y} \right)^2 \varphi^+_{{\scriptscriptstyle(1)},y} 
  	- \frac{\lambda}{2!} \varphi^+_{{\scriptscriptstyle(0)},y}\psi^{++}_{{\scriptscriptstyle(0)},yy}\right)\right.
  \\&\qquad\quad
  \left. 
  - D^{A-}_{xy} \left(j^-_{{\scriptscriptstyle(1)},y} - \frac{\lambda}{2!}\left(\varphi^-_{{\scriptscriptstyle(0)},y}\right)^2 \varphi^-_{{\scriptscriptstyle(1)},y}
  - \frac{\lambda}{2!} \varphi^-_{{\scriptscriptstyle(0)},y}\psi^{--}_{{\scriptscriptstyle(0)},yy}\right)\right] \,.
\end{aligned}\label{}
\end{equation}
%
Substituting the perturbiner expansions gives us,
\begin{equation}
\begin{aligned}
  \Phi^{\scriptscriptstyle{(1)}}_{p}\big|^{A}_{\mathcal{P}} 
  &= 
  -\frac{\lambda}{2!}D^{A+}_{\mathcal{P}}
  \left(
  	  \frac{1}{3}\left\lceil\Phi^{+},\Phi^{+},\Phi^{+}\right\rfloor^{{\scriptscriptstyle(1)}}_{p|\mathcal{P}} 
  	+  \int_p \left\lceil\Phi^{+},\Psi^{++}_{p}\right\rfloor^{{\scriptscriptstyle(0)}}_{\mathcal{P}}\right)
  \\&\quad
  +\frac{\lambda}{2!}D^{A-}_{\mathcal{P}} 
  \left(
  	  \frac{1}{3}\left\lceil\Phi^{-},\Phi^{-},\Phi^{-}\right\rfloor^{{\scriptscriptstyle(1)}}_{p|\mathcal{P}} 
  	+ \int_p \left\lceil\Phi^{-},\Psi^{--}_{p}\right\rfloor^{{\scriptscriptstyle(0)}}_{\mathcal{P}}\right)\,.
\end{aligned}\label{RecursionPhi1}
\end{equation}
The off-shell recursion for $\Psi^{{\scriptscriptstyle(0)}}\big|^{A,B}_{p|\mathcal{P}}$ can be derived by acting the descend operator $Q^{A}_{\ell}$ on \eqref{RecursionPhi0},
\begin{equation}
  \Psi^{{\scriptscriptstyle(0)}}\big|^{A,B}_{p|\mathcal{P}}
  =
  - \frac{\lambda}{3!} D^{A+}_{p|\mathcal{P}} Q^{B}_{p} \left\lceil\Phi^{+},\Phi^{+},\Phi^{+}\right\rfloor^{{\scriptscriptstyle(0)}}_{\mathcal{P}}
  + \frac{\lambda}{3!} D^{A-}_{p|\mathcal{P}} Q^{B}_{p} \left\lceil\Phi^{-},\Phi^{-},\Phi^{-}\right\rfloor^{{\scriptscriptstyle(0)}}_{\mathcal{P}}\,.
\label{RecursionPsi10}\end{equation}
The initial conditions for the one-loop off-shell currents are trivial \cite{Lee:2022aiu},
\begin{equation}
    \Phi^{{\scriptscriptstyle(1)}}_{p}\big|^{A}_i = 0\,.
\label{OneLoopInitialCurrent}\end{equation}
To derive the initial conditions for $\Psi^{{\scriptscriptstyle(0)}}\big|^{A,C}_{p|\emptyset}$, we examine the tree level terms in \eqref{1stDescDysonSchwinger},
\begin{equation}
\begin{aligned}
    \int_p \Psi^{{\scriptscriptstyle(0)}}\big|^{A,C}_{p|\emptyset} e^{ip\cdot(x-z)}
	= \int_p \tilde{D}^{AB}_{p} e^{ip\cdot(x-z)}\eta^{BC}\,,
\end{aligned}
\end{equation}
After comparing both sides above, we get,
\begin{gather}
    \Psi^{{\scriptscriptstyle(0)}}\big|^{A,C}_{p|\emptyset} = D^{AB}_{p|\emptyset}\eta^{BC}\,.
\label{Tree1stDescInitialCondition}\end{gather}
The initial conditions for higher order currents and descendants can be read off directly from the recursion relations using the initial conditions of previous currents. 

\subsubsection{Two-loop level}
To construct the two-loop level recursion relations, we substitute the $\hbar$ expansion \eqref{HbarExpansions_fields} into the DS equations \eqref{ExplicitDysonSchwinger} and collect the $\hbar^{2}$ order terms, 
\begin{equation}
\begin{aligned}
  &\varphi^A_{{\scriptscriptstyle(2)},x}
  = 
  \\&
  - \frac{\lambda}{2!} \int_{y}
  	D^{A+}_{xy}\left[
  		  \left(\varphi^+_{{\scriptscriptstyle(0)},y} \right)^2 \varphi^+_{{\scriptscriptstyle(2)},y} 
  		+ \left(\varphi^+_{{\scriptscriptstyle(1)},y} \right)^2 \varphi^+_{{\scriptscriptstyle(0)},y}
  		+ \varphi^+_{\scriptscriptstyle(0),y}\psi^{++}_{\scriptscriptstyle(1),yy}
  		+ \varphi^+_{\scriptscriptstyle(1),y}\psi^{++}_{\scriptscriptstyle(0),yy}
  	 	+ \frac{1}{3}\psi^{+++}_{\scriptscriptstyle(0),yyy}\right]
  \\&
  + \frac{\lambda}{2!} \int_{y} D^{A-}_{xy}\left[
  		  \left(\varphi^-_{{\scriptscriptstyle(0)},y} \right)^2 \varphi^-_{{\scriptscriptstyle(2)},y} 
  		+ \left(\varphi^-_{{\scriptscriptstyle(1)},y} \right)^2 \varphi^-_{{\scriptscriptstyle(0)},y} 
  		+ \varphi^-_{\scriptscriptstyle(0),y}\psi^{--}_{\scriptscriptstyle(1),yy} 
  		+ \varphi^-_{\scriptscriptstyle(1),y}\psi^{--}_{\scriptscriptstyle(0),yy}
  		+ \frac{1}{3}\psi^{---}_{\scriptscriptstyle(0),yyy} \right]
  \\&
  +\int_{y} \left[ D^{A+}_{xy} j^{+}_{{\scriptscriptstyle(2)},y} - D^{A-}_{xy} j^{-}_{{\scriptscriptstyle(2)},y}\right]\,.
\end{aligned}\label{}
\end{equation}
Substituting the perturbiner expansions gives us,
\begin{equation}
\begin{aligned}
  \Phi^{\scriptscriptstyle{(2)}}_{pq}\big|^{A}_{\mathcal{P}} 
  &= 
  -\frac{\lambda}{2!}D^{A+}_{\mathcal{P}}
  \left(
  	  \frac{1}{3}\left\lceil\Phi^{+},\Phi^{+},\Phi^{+}\right\rfloor^{{\scriptscriptstyle(2)}}_{pq|\mathcal{P}} 
  	+ \left\lceil\Phi^{+},\Psi^{++}_{p}\right\rfloor^{{\scriptscriptstyle(1)}}_{q|\mathcal{P}} + \frac13\Psi^{\scriptscriptstyle(0)}\big|^{+++}_{pq|\mathcal{P}} \right)
  \\&\quad
  +\frac{\lambda}{2!}D^{A-}_{\mathcal{P}} 
  \left(
  	  \frac{1}{3}\left\lceil\Phi^{-},\Phi^{-},\Phi^{-}\right\rfloor^{{\scriptscriptstyle(2)}}_{pq|\mathcal{P}} 
  	+ \left\lceil\Phi^{-},\Psi^{--}_{p}\right\rfloor^{{\scriptscriptstyle(1)}}_{q|\mathcal{P}}+ \frac13\Psi^{\scriptscriptstyle(0)}\big|^{---}_{pq|\mathcal{P}}\right)\,.
\end{aligned}\label{RecursionPhi2}
\end{equation}
The off-shell recursion for $\Psi^{{\scriptscriptstyle(1)}}_{p}\big|^{A,B}_{q|\mathcal{P}}$ can be derived by acting the descend operator on \eqref{RecursionPhi1},
\begin{equation}
\begin{aligned}
  \Psi^{\scriptscriptstyle{(1)}}_{p}\big|^{AB}_{q|\hat{\mathcal{P}}} 
  &= 
  -\frac{\lambda}{2!}D^{A+}_{q|\hat{\mathcal{P}}}Q^{B}_{q}
  \left(
  	  \frac{1}{3}\left\lceil\Phi^{+},\Phi^{+},\Phi^{+}\right\rfloor^{{\scriptscriptstyle(1)}}_{p|\hat{\mathcal{P}}} 
  	+ \left\lceil\Phi^{+},\Psi^{++}_{p}\right\rfloor^{{\scriptscriptstyle(0)}}_{\hat{\mathcal{P}}}\right)
  \\&\quad
  +\frac{\lambda}{2!}D^{A-}_{q|\hat{\mathcal{P}}}Q^{B}_{q}
  \left(
  	  \frac{1}{3}\left\lceil\Phi^{-},\Phi^{-},\Phi^{-}\right\rfloor^{{\scriptscriptstyle(1)}}_{p|\hat{\mathcal{P}}} 
  	+ \left\lceil\Phi^{-},\Psi^{--}_{p}\right\rfloor^{{\scriptscriptstyle(0)}}_{\hat{\mathcal{P}}}\right)\,,
\end{aligned}\label{RecursionPsi11}
\end{equation}
and for $\Psi^{\scriptscriptstyle(0)}\big|^{A,B,C}_{pq|\mathcal{P}}$, we derive the recursion relations similarly from \eqref{RecursionPsi10},
\begin{equation}
  \Psi^{{\scriptscriptstyle(0)}}\big|^{ABC}_{pq|\mathcal{P}}
  =
  - \frac{\lambda}{3!} D^{A+}_{pq|\mathcal{P}} Q^{BC}_{pq} \left\lceil\Phi^{+},\Phi^{+},\Phi^{+}\right\rfloor^{{\scriptscriptstyle(0)}}_{\mathcal{P}}
  + \frac{\lambda}{3!} D^{A-}_{pq|\mathcal{P}} Q^{BC}_{pq} \left\lceil\Phi^{-},\Phi^{-},\Phi^{-}\right\rfloor^{{\scriptscriptstyle(0)}}_{\mathcal{P}}\,.
\label{RecursionPsi20}\end{equation}
%
\subsubsection{Three-loop level}
To finally derive the three-loop level recursions, we follow the same procedure as earlier: first we substitute the $\hbar$ expansion \eqref{HbarExpansions_fields} into the DS equations \eqref{ExplicitDysonSchwinger}, then we collect the $\hbar^3$ order terms and substitute the perturbiner expansions into them. Here we skip explicitly listing the $\hbar^3$-order terms and directly showcase the recursion relations,
\begin{equation}
\begin{aligned}
  \Phi^{\scriptscriptstyle{(3)}}_{pqr}\big|^{A}_{\mathcal{P}} 
  &= 
  -\frac{\lambda}{2!}D^{A+}_{\mathcal{P}}
  \left(
  	  \frac{1}{3}\left\lceil\Phi^{+},\Phi^{+},\Phi^{+}\right\rfloor^{{\scriptscriptstyle(3)}}_{pqr|\mathcal{P}} 
  	+ \left\lceil\Phi^{+},\Psi^{++}_{p}\right\rfloor^{{\scriptscriptstyle(2)}}_{qr|\mathcal{P}} 
  	+ \frac13\Psi^{\scriptscriptstyle(1)}_{r}\big|^{+++}_{pq|\mathcal{P}} 
  \right)
  \\&\quad
  +\frac{\lambda}{2!}D^{A-}_{\mathcal{P}} 
  \left(
  	  \frac{1}{3}\left\lceil\Phi^{-},\Phi^{-},\Phi^{-}\right\rfloor^{{\scriptscriptstyle(3)}}_{pqr|\mathcal{P}} 
  	+ \left\lceil\Phi^{-},\Psi^{--}_{p}\right\rfloor^{{\scriptscriptstyle(2)}}_{qr|\mathcal{P}}
  	+ \frac{1}{3}\Psi^{\scriptscriptstyle(1)}_{r}\big|^{---}_{pq|\mathcal{P}}\right)\,.
\end{aligned}\label{RecursionPhi3}
\end{equation}
We obtain the off-shell recursion for the required descendant fields by acting the descend operator on \eqref{RecursionPhi2}, \eqref{RecursionPsi11} and \eqref{RecursionPsi20},
\begin{equation}
\begin{aligned}
  \Psi^{\scriptscriptstyle{(2)}}_{pq}\big|^{AB}_{r|\hat{\mathcal{P}}} 
  &= 
  -\frac{\lambda}{2!}D^{A+}_{r|\hat{\mathcal{P}}} Q^{B}_{r}
  \left(
  	  \frac{1}{3}\left\lceil\Phi^{+},\Phi^{+},\Phi^{+}\right\rfloor^{{\scriptscriptstyle(2)}}_{pq|\hat{\mathcal{P}}} 
  	+ \left\lceil\Phi^{+},\Psi^{++}_{p}\right\rfloor^{{\scriptscriptstyle(1)}}_{q|\hat{\mathcal{P}}} 
  	+ \frac{1}{3} \Psi^{\scriptscriptstyle(0)}\big|^{+++}_{pq|\hat{\mathcal{P}}} \right)
  \\&\quad
  +\frac{\lambda}{2!}D^{A-}_{r|\hat{\mathcal{P}}} Q^{B}_{r}
  \left(
  	  \frac{1}{3}\left\lceil\Phi^{-},\Phi^{-},\Phi^{-}\right\rfloor^{{\scriptscriptstyle(2)}}_{pq|\hat{\mathcal{P}}} 
  	+ \left\lceil\Phi^{-},\Psi^{--}_{p}\right\rfloor^{{\scriptscriptstyle(1)}}_{q|\hat{\mathcal{P}}}
  	+ \frac13\Psi^{\scriptscriptstyle(0)}\big|^{---}_{pq|\hat{\mathcal{P}}}\right)\,,
 \\
 \Psi^{\scriptscriptstyle{(1)}}_{p}\big|^{ABC}_{qr|\mathcal{P}} 
  &= 
  -\frac{\lambda}{2!}D^{A+}_{qr|\mathcal{P}} Q^{BC}_{qr}
  \left(
  	  \frac{1}{3}\left\lceil\Phi^{+},\Phi^{+},\Phi^{+}\right\rfloor^{{\scriptscriptstyle(1)}}_{p|\mathcal{P}} 
  	+ \left\lceil\Phi^{+},\Psi^{++}_{p}\right\rfloor^{{\scriptscriptstyle(0)}}_{\mathcal{P}}\right)
  \\&\quad
  +\frac{\lambda}{2!}D^{A-}_{qr|\mathcal{P}} Q^{BC}_{qr}
  \left(
  	  \frac{1}{3}\left\lceil\Phi^{-},\Phi^{-},\Phi^{-}\right\rfloor^{{\scriptscriptstyle(1)}}_{p|\mathcal{P}} 
  	+ \left\lceil\Phi^{-},\Psi^{--}_{p}\right\rfloor^{{\scriptscriptstyle(0)}}_{\mathcal{P}}\right)\,,
  \\
  \Psi^{{\scriptscriptstyle(0)}}\big|^{ABCD}_{pqr|\hat{\mathcal{P}}}
  &=
  - \frac{\lambda}{3!} D^{A+}_{pqr|\hat{\mathcal{P}}} Q^{BCD}_{pqr} \left\lceil\Phi^{+},\Phi^{+},\Phi^{+}\right\rfloor^{{\scriptscriptstyle(0)}}_{\hat{\mathcal{P}}}
  + \frac{\lambda}{3!} D^{A-}_{pqr|\hat{\mathcal{P}}} Q^{BCD}_{pqr} \left\lceil\Phi^{-},\Phi^{-},\Phi^{-}\right\rfloor^{{\scriptscriptstyle(0)}}_{\hat{\mathcal{P}}}\,.
\end{aligned}\label{RecursionPsi30}
\end{equation}

We have derived the recursion relations and generated the off-shell currents up to the three-loop order. As previously discussed, the off-shell currents (and their descendants) are the amputated correlation functions that include one or more off-shell legs.

\subsection{Filtering the relevant amplitudes}
In this work, we focus on $2 \to n$ tree-level differential cross-sections, $\mathrm{d} \sigma_{2\to n} = (\mathcal{M}^{{\scriptscriptstyle(0)}}_{n+2})^{*} \mathcal{M}^{{\scriptscriptstyle(0)}}_{n+2}$. According to the optical theorem, $\mathrm{d}\sigma_{2\to n}$ can be derived by considering $2 \to 2$ amplitude in the doubling prescription, $\mathcal{A}_{4}$. However, $\mathcal{A}_{4}$ includes different cross-sections with different loop orders. Schematically, $\mathcal{A}_{4}$ is related with $\mathcal{M}_{n}$ as follows:
\begin{equation}
\begin{aligned}
  \lambda^{2} \mathcal{A}^{{\scriptscriptstyle(1)}}_{4} 
  &\sim
  \lambda^{2} \big(\mathcal{M}^{{\scriptscriptstyle(0)}}_{4}\big)^{*} \mathcal{M}^{{\scriptscriptstyle(0)}}_{4}
  \\
  \lambda^{3} \mathcal{A}^{{\scriptscriptstyle(2)}}_{4} 
  &\sim 
    \lambda^{3} \big(\mathcal{M}^{{\scriptscriptstyle(1)}}_{4}\big)^{*} \mathcal{M}^{{\scriptscriptstyle(0)}}_{4} 
  + \lambda^{3} \big(\mathcal{M}^{{\scriptscriptstyle(0)}}_{4}\big)^{*} \mathcal{M}^{{\scriptscriptstyle(1)}}_{4}
  \\
  \lambda^{4} \mathcal{A}^{{\scriptscriptstyle(3)}}_{4} 
  &\sim 
    \lambda^{4} \big(\mathcal{M}^{{\scriptscriptstyle(0)}}_{6}\big)^{*} \mathcal{M}^{{\scriptscriptstyle(0)}}_{6} 
  + \lambda^{4} \big(\mathcal{M}^{{\scriptscriptstyle(1)}}_{4}\big)^{*} \mathcal{M}^{{\scriptscriptstyle(1)}}_{4}
  + \lambda^{4} \big(\mathcal{M}^{{\scriptscriptstyle(2)}}_{4}\big)^{*} \mathcal{M}^{{\scriptscriptstyle(0)}}_{4}
  + \lambda^{4} \big(\mathcal{M}^{{\scriptscriptstyle(0)}}_{4}\big)^{*} \mathcal{M}^{{\scriptscriptstyle(2)}}_{4} \,.
\end{aligned}\label{}
\end{equation}
Here, we explicitly pulled out the coupling constant to see the manifest dependence of the order of the couplings.
This can be easily checked through the following power counting of $\mathcal{M}$ with respect to the coupling constant $\lambda$
\begin{equation}
\begin{aligned}
    \mathcal{M}_4 &= \lambda \mathcal{M}_4^{(0)} + \lambda^2 \mathcal{M}_4^{(1)} + \lambda^3 \mathcal{M}_4^{(2)} + \lambda^4 \mathcal{M}_4^{(3)} + \mathcal{O}(\lambda^5)\,,
    \\
	\mathcal{M}_6 &= \lambda^2 \mathcal{M}_6^{(0)} + \lambda^3 \mathcal{M}_6^{(1)} + \lambda^4 \mathcal{M}_6^{(2)} + \mathcal{O}(\lambda^5)\,.
\end{aligned}
\end{equation}

This means that if we compute the 3-loop doubled amplitude $\mathcal{A}^{{\scriptscriptstyle(3)}}_{4}$ to obtain tree-level $2\to 4$ differential cross-section $\mathrm{d}\sigma^{{\scriptscriptstyle(0)}}_{2\to 4}$, we must eliminate the unnecessary contributions from the two-loop $2\to 2$ differential cross-section $\mathrm{d}\sigma^{{\scriptscriptstyle(2)}}_{2\to 2}$. For instance, diagram (a) in figure \ref{fig:WantedvUnwanted} is one of the diagrams contributing to $\mathrm{d}\sigma^{{\scriptscriptstyle(0)}}_{2\to 4}$. On the other hand, diagram (b) in figure \ref{fig:WantedvUnwanted} contributes to $\mathrm{d}\sigma^{{\scriptscriptstyle(2)}}_{2\to 2}$. Therefore, we derive the filtering process that allows us to extract only the required cross-section from the entire doubled amplitude.

After imposing cutting rules defined in \eqref{Def_cutting}, the original diagram is decomposed into sub-diagrams consisting entirely of either circled or uncircled vertices. We categorize these sub-diagrams into two sectors: diagrams consisting of only circled vertices are the ``$+$ sector'', while those consisting of uncircled vertices are the ``$-$ sector''. Each sector contains at least one diagram, but we require that the number of diagrams in each sector has to be one because we need connected diagrams only. 

To describe the filtering process clearly, we introduce the following notations:
\begin{itemize}
  \item $N^{AB}$ : Number of $D^{AB}$ propagators. For instance, $N^{++}$ and $N^{+-}$ are numbers of $D^{++}$ and $D^{+-}$ propagators, or $D_{F}$ and $\Delta^{+}$, respectively. 
  \item $V^{\pm}$ : Numbers of vertices in $\pm$ sector.
  \item $L_{0}$ : Number of total loops before the cutting.
  \item $L$ : Number of total loops in both $+$ and $-$ sectors.
  \item $C$ : Number of total disconnected diagrams in both $+$ and $-$ sectors.
\end{itemize}
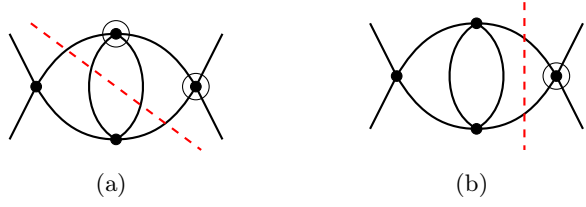
\begin{figure}[t]
\begin{center}
\subfigure[]{
 \begin{tikzpicture}[scale=1.4]
    \draw[thick] (-0.75,0) -- (-1,-0.5);
    \draw[thick] (-0.75,0) -- (-1,0.5);
    \draw[thick] (-0.75,0) to[out=60, in =180] (0,0.5);
    \draw[thick] (0,0.5) to[out=-150, in =150] (0,-0.5);
    \draw[thick] (0,0.5) to[out=-30, in =30] (0,-0.5);
    \draw[thick] (0.75,0) to[out=120, in =0] (0,0.5);
    \draw[thick] (0.75,0) to[out=-120, in =0] (0,-0.5);
    \draw[thick] (-0.75,0) to[out=-60, in =180] (0,-0.5);
    \draw[thick] (0.75,0) -- (1,-0.5);
    \draw[thick] (0.75,0) -- (1,0.5);
    \draw[fill=black] (-0.75,0) circle (0.05);
    \draw[fill=black] (0,0.5) circle (0.05);
    \draw (0,0.5) circle (0.125);
    \draw[fill=black] (0.75,0) circle (0.05);
    \draw (0.75,0) circle (0.125);
    \draw[fill=black] (0,-0.5) circle (0.05);
    \draw[thick,dashed,red] (-0.8,0.6) -- (0.8, -0.6);
\end{tikzpicture}
}
\qquad\qquad
\subfigure[]{
 \begin{tikzpicture}[scale=1.4]
    \draw[thick] (-0.75,0) -- (-1,-0.5);
    \draw[thick] (-0.75,0) -- (-1,0.5);
    \draw[thick] (-0.75,0) to[out=60, in =180] (0,0.5);
    \draw[thick] (0,0.5) to[out=-150, in =150] (0,-0.5);
    \draw[thick] (0,0.5) to[out=-30, in =30] (0,-0.5);
    \draw[thick] (0.75,0) to[out=120, in =0] (0,0.5);
    \draw[thick] (0.75,0) to[out=-120, in =0] (0,-0.5);
    \draw[thick] (-0.75,0) to[out=-60, in =180] (0,-0.5);
    \draw[thick] (0.75,0) -- (1,-0.5);
    \draw[thick] (0.75,0) -- (1,0.5);
    \draw[fill=black] (-0.75,0) circle (0.05);
    \draw[fill=black] (0,0.5) circle (0.05);
    \draw[fill=black] (0.75,0) circle (0.05);
    \draw (0.75,0) circle (0.125);
    \draw[fill=black] (0,-0.5) circle (0.05);
    \draw[thick,dashed,red] (0.45,0.7) -- (0.45, -0.7);
\end{tikzpicture}
}
\end{center}
\vspace{-\baselineskip}
\caption{Two diagrams that contribute to $\mathcal{A}^{{\scriptscriptstyle(3)}}_{1^{+}2^{-}3^{-}4^{+}}$. (a) is relevant to $\sigma^{{\scriptscriptstyle(0)}}_{2\to 4}$, and (b) is relevant to $\sigma^{{\scriptscriptstyle(2)}}_{2\to 2}$. Since we are considering only the tree-level cross-section, we devise a method for selecting terms like (a) in an algebraic way.} 
\label{fig:WantedvUnwanted}
\end{figure}

Note that the number of cuts corresponding to the cross-section for $2 \to n$ scattering should be $n$. Thus we have the following relation
\begin{equation}
  N^{+-} + N^{-+} = n \,.
\label{}\end{equation}
In addition, after the cutting, it should satisfy the well-known relation among the number of loops, vertices and propagators,
\begin{equation}
  C = L - N^{++} - N^{--} + V^{+} + V^{-}\,.
\label{Euler_formula}\end{equation}
%
Since we are only interested in connected diagrams, both $C^{+}$ and $C^{-}$ must be equal to 1. 

We next determine $L$ from the information of propagators. Let us extract only the loop momentum part from $\Delta^{\pm}$ and denote them as $v_{i}$, where $i$ is a labeling of $\Delta^{\pm}$. We define a set of all $v_{i}$ as $V = \{v_{1},\cdots v_{n}\}$. For instance, in $\Delta^{\pm}_{p}$ and $\Delta^{\pm}_{pq|23}$, the relevant vectors correspond to $p$ and $p+q$, respectively. Let $l$ denotes the number of linearly independent vectors in the entire set $V$. If we denote the loop order before the cutting as $L_{0}$ number of loop momenta is $n-1$, the remaining number of loops after cutting is given by  
\begin{equation}
  L=L_{0}-l
\label{}\end{equation}
The simplest way to determine the value of $l$ is by calculating the rank of the following matrix $M$
\begin{equation}
  V = M P\,,
  \qquad
  l = \mbox{Rank}~ M
\label{}\end{equation}
where $P$ is the vector of collection of all the loop momenta
\begin{equation}
  P = \{p_{1},p_{2},\cdots, p_{n-1}\}\,.
\label{}\end{equation}

Let us consider a specific example in figure \ref{fig:WantedvUnwanted}.
\begin{itemize}
  \item{Diagram (a):}\\ 
  This diagram is represented by $\tilde{D}^{+}_{p|23}
    \tilde{\Delta}^{+}_{pqr|23} \tilde{\Delta}^{-}_{p|\emptyset} \tilde{\Delta}^{-}_{q|\emptyset} \tilde{\Delta}^{-}_{r|\emptyset} \tilde{D}^{-}_{q|23}$. The corresponding $V$ and $P$ vectors are $V=\{p,q,r,pqr\}$ and $P=\{p,q,r\}$. Then the matrix $M$ and its rank are given by
\begin{equation}
  M = \begin{pmatrix}
        1 & 0 & 0 \\
        0 & 1 & 0 \\
        0 & 0 & 1 \\
        1 & 1 & 1
    \end{pmatrix} \,,
    \qquad
    l=\mbox{rank}~ M = 3\,.
\label{}\end{equation}
Since $n=N^{+-} + N^{-+} =4$, this diagram corresponds to $2\to4$ scattering. The total loop order after the cutting is $L=3-3 = 0$. Then $C = 0 - 1-1 +2+2 = 2$, and this means after the cutting, the diagram is separated into 1 circled and 1 uncircled sub-diagrams. Thus it corresponds to $\sigma^{\ord{0}}_{2\to4}$.
\item{Diagram (b):}\\
	In this case, the diagram represented by $\tilde{D}^{+}_{q|\emptyset} \tilde{D}^{+}_{r|\emptyset} \tilde{D}^{+}_{r|23} \tilde{D}^{+}_{pqr|23} \tilde{\Delta}^{+}_{p|23} \tilde{\Delta}^{-}_{p|\emptyset}$. In this case, the $V$ and $P$ vectors are $V=\{p\}$ and $P=\{p,q,r\}$. The $M$ matrix and its rank are given by
\begin{equation}
  M = 
  \begin{pmatrix} 
  1 & 0 & 0
  \end{pmatrix}
  \,,
  \qquad
  l = \mbox{rank}~ M = 1
\label{}\end{equation}
Since $n = N^{+-} + N^{-+} =2$, this diagram corresponds to $2\to 2$ scattering. The total loop order after the cutting is $L = 3 - 1 = 2$. is also divided into one circled and one uncircled sub-diagrams. Then $C = 2 - 4- 0 + 3 + 1 = 2$, and this means after the cutting, the diagram is separated into 1 circled and 1 uncircled sub-diagrams. The circled one is tree level, satisfying $0= 0-1+1$. Thus it corresponds to $\sigma^{\ord{2}}_{2\to2}$. 
\end{itemize}

The following is an algorithm for algebraic filtering that does not rely on diagrams to extract only the loop integrands that contribute to $L$-loop $2\to n$ cross-section $\sigma^{\ord{L}}_{2\to n}$:
\begin{enumerate}
\item[(1)] Choose the values $L_{0}$ and $n$. 
\item[(2)] Collect the terms satisfying $N^{+-} + N^{-+} =n$.
\item[(3)] Compute $L$ and $C$ values for each integrand. Keep the terms only $C=2$.
\item[(4)] Remove terms that do not satisfy the relations \eqref{Euler_formula}. 
\end{enumerate}
After this process, the remaining terms contribute to $\sigma^{\ord{L}}_{2\to n}$ only.

\section{Differential cross-section from the recursions}\label{Sec:4}
So far, we have derived the recursion relations from the Dyson--Schwinger equations and the algorithm for filtering out the irrelevant terms. Here, we present the doubled amplitudes $\mathcal{A}_{1^+, 2^-, 3^-, 4^+}$ by solving the recursion relations up to three-loop orders. We will truncate all terms except those corresponding to the $2\to n$ tree-level differential cross-section, $\mathrm{d}\sigma^{{\scriptscriptstyle(0)}}_{2\to n}$, using the filtering.

\subsection{Filtered off-shell currents up to three-loop level}

At the tree level, the solution of the recursion is trivial. It is straightforward to see that there are no cut diagrams with two $+$ legs and two $-$ legs.
\begin{align}
    \mathcal{A}^{{\scriptscriptstyle(0)}}_{1^+2^- 3^-4^+} = 0 \,.
\end{align}
At one-loop level, the off-shell current $\Phi^{{\scriptscriptstyle(1)}}_{1^{+}2^{-}3^{-}}$ is easily derived by solving the recursion. It gives $\mathrm{d} \sigma^{{\scriptscriptstyle(0)}}_{2\to 2}$
\begin{align}
    \mathcal{A}^{{\scriptscriptstyle(1)}}_{1^+2^-3^-4^+} = \frac{\lambda^2}{2} \int_p \tilde{\Delta}^{+}_{p|23} \, \tilde{\Delta}^{-}_p \,.
\label{one_loop_doubled_amp}\end{align}
At two-loop level, $\Phi^{{\scriptscriptstyle(2)}}_{1^{+}2^{-}3^{-}}$ contains only terms corresponding to $\mathrm{d}\sigma^{{\scriptscriptstyle(1)}}_{2\to 2}$, and it is irrelevant to our interest
\begin{equation}
  \mathcal{A}^{{\scriptscriptstyle(2)}}_{1^+2^-3^-4^+} \sim 0 \,,
\label{}\end{equation}
where $\sim$ represents equality after excluding terms that do not contribute to $\mathrm{d}\sigma^{{\scriptscriptstyle(0)}}_{2\to n}$.

\begin{figure}[t]
    \centering
\subfigure[]{
\begin{tikzpicture}
     \node[] at (-1,0.5) [left] {1};
     \node[] at (-1,-0.5) [left] {4};
     \draw[thick] (-0.75,0) -- (-1,-0.5);
    \draw[thick] (-0.75,0) -- (-1,0.5);
    \draw[thick] (-0.75,0) to[out=60, in =180] (0,0.5);
    \draw[thick] (0,0.5) to[out=-150, in =150] (0,-0.5);
     \draw[thick] (0,0.5) to[out=-30, in =30] (0,-0.5);
    \draw[thick] (0.75,0) to[out=120, in =0] (0,0.5);
    \draw[thick] (0.75,0) to[out=-120, in =0] (0,-0.5);
    \draw[thick] (-0.75,0) to[out=-60, in =180] (0,-0.5);
    \draw[thick] (0.75,0) -- (1,-0.5);
    \draw[thick] (0.75,0) -- (1,0.5);
    \draw[fill=black] (-0.75,0) circle (0.05);
    \draw[fill=black] (0,0.5) circle (0.05);
    \draw (0,0.5) circle (0.125);
    \draw[fill=black] (0.75,0) circle (0.05);
    \draw (0.75,0) circle (0.125);
    \draw[fill=black] (0,-0.5) circle (0.05);
    \node[] at (1,0.5) [right] {2};
    \node[] at (1,-0.5) [right] {3};
    \draw[thick,dashed,red] (-0.8,0.6) -- (0.8, -0.6);
\end{tikzpicture}\label{fig:type1}}
\qquad
\subfigure[]{
\begin{tikzpicture}
    \node[] at (-1,0.5) [left] {1};
    \node[] at (-1,-0.5) [left] {4};
    \node[] at (1,0.5) [right] {2};
    \node[] at (1,-0.5) [right] {3};
    \draw[fill=black] (-0.75,0) circle (0.05);
    \draw[fill=black] (0.75,0) circle (0.05);
    \draw (0.75,0) circle (0.125);
    \draw[thick] (-1,0.5) -- (-0.75,0);
    \draw[thick] (-1,-0.5) -- (-0.75,0);
    \draw[thick] (1,0.5) -- (0.75,0);
    \draw[thick] (1,-0.5) -- (0.75,0);
    \draw[thick] (-0.75,0) to[out=70, in=110] (0.75,0);
    \draw[thick] (-0.75,0) to[out=-70, in=-110] (0.75,0);
    \draw[fill=black] (-0.375,0.35) circle (0.05);
    \draw[fill=black] (0.375,0.35) circle (0.05);
    \draw (0.375,0.35) circle (0.125);
    \draw[thick] (-0.375,0.35) to[out=70, in=110] (0.375,0.35);
    \draw[thick] (-0.375,0.35) to[out=-70, in=-110] (0.375,0.35);
    \draw[thick,dashed,red] (0,0.7) -- (0,-0.6);
\end{tikzpicture}\label{fig:type2}}
\qquad
\subfigure[]{
\begin{tikzpicture}
     \node[] at (-1,0.5) [left] {1};
     \node[] at (-1,-0.5) [left] {4};
     \draw[thick] (-0.75,-0.3) -- (-1,-0.5);
    \draw[thick] (-0.75,0.3) -- (-1,0.5);
    \draw[thick] (-0.75,-0.3) to[out=120, in =-120] (-0.75,0.3);
    \draw[thick] (0,0.5) to[out=-150, in =0] (-0.75,0.3);
    \draw[thick] (0,0.5) to[out=-120, in =30] (-0.75,-0.3);
    \draw[thick] (-0.75,0.3) to[out=60, in =150] (0,0.5);
    \draw[thick] (0.75,0) to[out=-120, in =-30] (-0.75,-0.3);
    \draw[thick] (0,0.5) to[out=-10, in =130] (0.75,0);
    \draw[thick] (0.75,0) -- (1,-0.5);
    \draw[thick] (0.75,0) -- (1,0.5);
    \draw[fill=black] (-0.75,0.3) circle (0.05);
    \draw[fill=black] (0,0.5) circle (0.05);
    \draw (0,0.5) circle (0.125);
    \draw[fill=black] (0.75,0) circle (0.05);
    \draw (0.75,0) circle (0.125);
    \draw[fill=black] (-0.75,-0.3) circle (0.05);
    \node[] at (1,0.5) [right] {2};
    \node[] at (1,-0.5) [right] {3};
    \draw[thick,dashed,red] (-0.5,0.7) to[out=-90, in=160] (0.6, -0.6);
\end{tikzpicture}\label{fig:type3}}
\qquad
\subfigure[]{
\begin{tikzpicture}
     \node[] at (-1,0.5) [left] {4};
     \node[] at (-1,-0.5) [left] {1};
     \draw[thick] (-0.75,-0.3) -- (-1,-0.5);
    \draw[thick] (-0.75,0.3) -- (-1,0.5);
    \draw[thick] (-0.75,-0.3) to[out=120, in =-120] (-0.75,0.3);
    \draw[thick] (0,0.5) to[out=-150, in =0] (-0.75,0.3);
    \draw[thick] (0,0.5) to[out=-120, in =30] (-0.75,-0.3);
    \draw[thick] (-0.75,0.3) to[out=60, in =150] (0,0.5);
    \draw[thick] (0.75,0) to[out=-120, in =-30] (-0.75,-0.3);
    \draw[thick] (0,0.5) to[out=-10, in =130] (0.75,0);
    \draw[thick] (0.75,0) -- (1,-0.5);
    \draw[thick] (0.75,0) -- (1,0.5);
    \draw[fill=black] (-0.75,0.3) circle (0.05);
    \draw[fill=black] (0,0.5) circle (0.05);
    \draw (0,0.5) circle (0.125);
    \draw[fill=black] (0.75,0) circle (0.05);
    \draw (0.75,0) circle (0.125);
    \draw[fill=black] (-0.75,-0.3) circle (0.05);
    \node[] at (1,0.5) [right] {2};
    \node[] at (1,-0.5) [right] {3};
    \draw[thick,dashed,red] (-0.5,0.7) to[out=-90, in=160] (0.6, -0.6);
\end{tikzpicture}\label{fig:type4}}\\
\subfigure[]{
\begin{tikzpicture}
    \node[] at (-1,0.5) [left] {1};
    \node[] at (-1,-0.5) [left] {4};
    \node[] at (1,0.5) [right] {2};
    \node[] at (1,-0.5) [right] {3};
    \draw[thick] (-0.75,0) -- (-1,0.5);
    \draw[thick] (-0.75,0) -- (-1,-0.5);
    \draw[fill=black] (-0.75,0) circle (0.05); 
    \draw[thick] (0.75,-0.3) -- (1,-0.5);
    \draw[thick] (0.75,0.3) -- (1,0.5);
    \draw[fill=black] (0.75,0.3) circle (0.05);
    \draw[fill=black] (0.75,-0.3) circle (0.05);
    \draw (0.75,-0.3) circle (0.125);
    \draw (0.75,0.3) circle (0.125);
    \draw[thick] (0.75,-0.3) to[out=60, in=-60] (0.75,0.3);
    \draw[fill=black] (0,0.5) circle (0.05);
    \draw[thick] (0,0.5) to[out=-30, in =180] (0.75,0.3);
    \draw[thick] (0,0.5) to[out=-60, in =150] (0.75,-0.3);
    \draw[thick] (0.75,0.3) to[out=120, in =30] (0,0.5);
    \draw[thick] (-0.75,0) to[out=-60, in=-150] (0.75,-0.3);
    \draw[thick] (-0.75,0) to[out=50, in=-170] (0,0.5);
    \draw[thick,dashed,red] (0.5,0.7) to[out=-90, in=20] (-0.6, -0.6);
\end{tikzpicture}\label{fig:type5}}
\qquad
\subfigure[]{
\begin{tikzpicture}
    \node[] at (-1,0.5) [left] {1};
    \node[] at (-1,-0.5) [left] {4};
    \node[] at (1,0.5) [right] {3};
    \node[] at (1,-0.5) [right] {2};
    \draw[thick] (-0.75,0) -- (-1,0.5);
    \draw[thick] (-0.75,0) -- (-1,-0.5);
    \draw[fill=black] (-0.75,0) circle (0.05); 
    \draw[thick] (0.75,-0.3) -- (1,-0.5);
    \draw[thick] (0.75,0.3) -- (1,0.5);
    \draw[fill=black] (0.75,0.3) circle (0.05);
    \draw[fill=black] (0.75,-0.3) circle (0.05);
    \draw (0.75,-0.3) circle (0.125);
    \draw (0.75,0.3) circle (0.125);
    \draw[thick] (0.75,-0.3) to[out=60, in=-60] (0.75,0.3);
    \draw[fill=black] (0,0.5) circle (0.05);
    \draw[thick] (0,0.5) to[out=-30, in =180] (0.75,0.3);
    \draw[thick] (0,0.5) to[out=-60, in =150] (0.75,-0.3);
    \draw[thick] (0.75,0.3) to[out=120, in =30] (0,0.5);
    \draw[thick] (-0.75,0) to[out=-60, in=-150] (0.75,-0.3);
    \draw[thick] (-0.75,0) to[out=50, in=-170] (0,0.5);
    \draw[thick,dashed,red] (0.5,0.7) to[out=-90, in=20] (-0.6, -0.6);
\end{tikzpicture}\label{fig:type6}}
\qquad
\subfigure[]{
\begin{tikzpicture}
    \node[] at (-1,0.5) [left] {1};
    \node[] at (-1,-0.5) [left] {4};
    \node[] at (1,0.5) [right] {2};
    \node[] at (1,-0.5) [right] {3};
    \draw[fill=black] (0.75,0.3) circle (0.05);
    \draw[fill=black] (0.75,-0.3) circle (0.05);
    \draw[fill=black] (-0.75,0.3) circle (0.05);
    \draw[fill=black] (-0.75,-0.3) circle (0.05);
    \draw (0.75,0.3) circle (0.125);
    \draw (0.75,-0.3) circle (0.125);
    \draw[thick] (-1,0.5) -- (-0.75, 0.3);
    \draw[thick] (-1,-0.5) -- (-0.75, -0.3);
    \draw[thick] (1,0.5) -- (0.75, 0.3);
    \draw[thick] (1,-0.5) -- (0.75, -0.3);
    \draw[thick] (0.75,0.3) -- (0.75, -0.3);
    \draw[thick] (-0.75,-0.3) -- (-0.75, 0.3);
    \draw[thick] (-0.75,0.3) to[out=15, in=165] (0.75,0.3);
    \draw[thick] (-0.75,0.3) to[out=-15, in=-165] (0.75,0.3);
    \draw[thick] (-0.75,-0.3) to[out=15, in=165] (0.75,-0.3);
    \draw[thick] (-0.75,-0.3) to[out=-15, in=-165] (0.75,-0.3);
    \draw[thick,dashed,red] (0,0.65) -- (0,-0.65);
\end{tikzpicture}\label{fig:type7}}
\qquad
\subfigure[]{
\begin{tikzpicture}
    \node[] at (-1,0.5) [left] {4};
    \node[] at (-1,-0.5) [left] {1};
    \node[] at (1,0.5) [right] {2};
    \node[] at (1,-0.5) [right] {3};
    \draw[fill=black] (0.75,0.3) circle (0.05);
    \draw[fill=black] (0.75,-0.3) circle (0.05);
    \draw[fill=black] (-0.75,0.3) circle (0.05);
    \draw[fill=black] (-0.75,-0.3) circle (0.05);
    \draw (0.75,0.3) circle (0.125);
    \draw (0.75,-0.3) circle (0.125);
    \draw[thick] (-1,0.5) -- (-0.75, 0.3);
    \draw[thick] (-1,-0.5) -- (-0.75, -0.3);
    \draw[thick] (1,0.5) -- (0.75, 0.3);
    \draw[thick] (1,-0.5) -- (0.75, -0.3);
    \draw[thick] (0.75,0.3) -- (0.75, -0.3);
    \draw[thick] (-0.75,-0.3) -- (-0.75, 0.3);
    \draw[thick] (-0.75,0.3) to[out=15, in=165] (0.75,0.3);
    \draw[thick] (-0.75,0.3) to[out=-15, in=-165] (0.75,0.3);
    \draw[thick] (-0.75,-0.3) to[out=15, in=165] (0.75,-0.3);
    \draw[thick] (-0.75,-0.3) to[out=-15, in=-165] (0.75,-0.3);
    \draw[thick,dashed,red] (0,0.65) -- (0,-0.65);
\end{tikzpicture}\label{fig:type8}}\\
\subfigure[]{
\begin{tikzpicture}
    \node[] at (-1,0.75) [left] {1};
    \node[] at (-1,-0.75) [left] {4};
    \draw[fill=black] (-0.75,0.4) circle (0.05);
    \draw[fill=black] (-0.75,-0.4) circle (0.05);
    \draw[fill=black] (-0.05,-0.06) circle (0.05);
    \draw (-0.05, -0.06) circle (0.125);
    \draw[fill=black] (0.15,0.72) circle (0.05);
    \draw (0.15,0.72) circle (0.125);
    \node[] at (0.6,0.9) [right] {2};
    \node[] at (0.42,-0.2) [right] {3};
    \draw[thick, dashed] (-0.75,0.4) -- (0.15,0.72);
    \draw[thick, dashed] (-0.75,-0.4) -- (0.15,0.72);
    \draw[thick, dashed] (-0.05,-0.06) -- (0.15,0.72);
    \draw[thick] (-0.75,0.4) -- (-0.75,-0.4) -- (-0.05,-0.06) --cycle;
    \draw[thick] (-1,0.75) -- (-0.75,0.4);
    \draw[thick] (-1,-0.75) -- (-0.75,-0.4);
    \draw[thick] (0.42,-0.2) -- (-0.05,-0.06);
    \draw[thick] (0.6, 0.9) -- (0.15,0.72);
    \draw[thick,dashed,red] (-0.2, 0.9) -- (-0.5, -0.75);
\end{tikzpicture}\label{fig:type9}}
\vspace{-0.5\baselineskip}
\caption{We present all the cut diagrams in total nine, which contribute to $\mathrm{d}\sigma^{{\scriptscriptstyle(0)}}_{2\to 4}$}
\label{fig:NineDiagrams}
\end{figure}
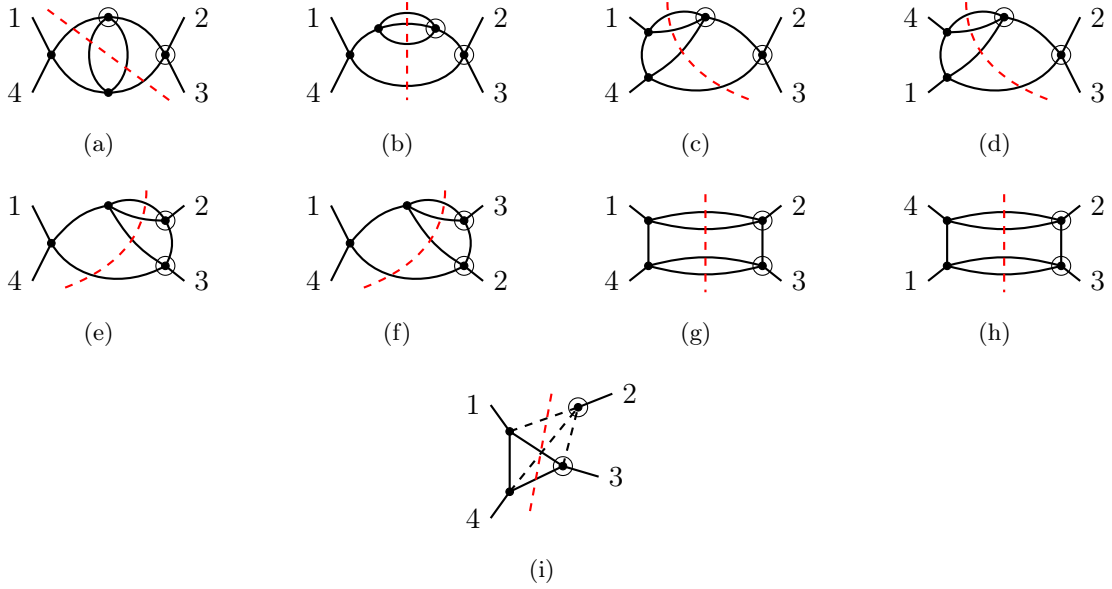

At three-loop level, $\Phi^{{\scriptscriptstyle(2)}}_{1^{+}2^{-}3^{-}}$ has only nine relevant terms to $\mathrm{d}\sigma^{{\scriptscriptstyle(0)}}_{2\to 4}$ after the filtering process,
\begin{align}
\begin{split}
  \mathcal{A}^{{\scriptscriptstyle(3)}}_{1^+ 2^- 3^- 4^+} 
  &\sim 
  \lambda^4 \int_{p,q,r} \Bigg[ 
    \frac{1}{2} \tilde{D}^{+}_{p|23}
    \tilde{\Delta}^{+}_{pqr|23}
    \tilde{\Delta}^{-}_{p}
    \tilde{\Delta}^{-}_{q}
    \tilde{\Delta}^{-}_{r}
    \tilde{D}^{-}_{q|23}
  \\&\qquad
  + \frac{1}{6} \tilde{D}^{+}_{p|23}
  	\tilde{\Delta}^{+}_{pqr|23}
  	\tilde{\Delta}^{-}_p
  	\tilde{\Delta}^{-}_q
  	\tilde{\Delta}^{-}_r 
  	\tilde{D}^{-}_{p|23} 
  + \frac{1}{2} \tilde{D}^{+}_{pq|123}
  	\tilde{\Delta}^{+}_{pqr|23} 
  	\tilde{\Delta}^{-}_p
  	\tilde{\Delta}^{-}_q
  	\tilde{\Delta}^{-}_r 
  	\tilde{D}^{-}_{p|23}
  \\& \qquad
  + \frac{1}{2}\tilde{D}^{+}_{pq|1} 
  	\tilde{\Delta}^{+}_{pqr|23}
  	\tilde{\Delta}^{-}_{p}
  	\tilde{\Delta}^{-}_q
  	\tilde{\Delta}^{-}_r
  	\tilde{D}^{-}_{p|23} 
  + \frac{1}{2}\tilde{D}^{+}_{p|23} 
  	\tilde{\Delta}^{+}_{pqr|23}
  	\tilde{\Delta}^{-}_p
  	\tilde{\Delta}^{-}_q
  	\tilde{\Delta}^{-}_r 
  	\tilde{D}^{-}_{pq|3}
    \\&\qquad
  + \frac{1}{2}\tilde{D}^{+}_{p|23} 
  	\tilde{\Delta}^{+}_{pqr|23}
  	\tilde{\Delta}^{-}_{p}
  	\tilde{\Delta}^{-}_{q}
  	\tilde{\Delta}^{-}_{r}
  	\tilde{D}^{+}_{pq|2} 
  + \frac{1}{4}\tilde{D}^{+}_{pq|1}
  	\tilde{\Delta}^{+}_{pqr|23}
  	\tilde{\Delta}^{-}_{p}
  	\tilde{\Delta}^{-}_{q}
  	\tilde{\Delta}^{-}_{r}
  	\tilde{D}^{-}_{pq|2} 
    \\&\qquad
  + \frac{1}{4}\tilde{D}^{+}_{pq|1}
  	\tilde{\Delta}^{+}_{pqr|23}
  	\tilde{\Delta}^{-}_{p}
  	\tilde{\Delta}^{-}_{q}
  	\tilde{\Delta}^{-}_{r}
  	\tilde{D}^{-}_{pq|3} 
  + \tilde{D}^{+}_{pq|123}
  	\tilde{\Delta}^{+}_{pqr|23}
  	\tilde{\Delta}^{-}_{p}
  	\tilde{\Delta}^{-}_{q}
  	\tilde{\Delta}^{-}_{r}
  	\tilde{D}^{-}_{pr|2} \Bigg] \,.
\end{split}\label{three_loop_doubled_amp}
\end{align}
We may identify each term in the above to the following diagrams in figure \ref{fig:NineDiagrams}.

As we have seen so far, these results are obtained directly by solving the off-shell recursions instead of squaring the amplitude after computing it. This approach is much more efficient than the conventional method of squaring the amplitude, and it will be a highly efficient tool for calculating cross-sections in more complicated theories, such as QCD or the Standard Model.

\subsection{Comparing with the differential cross-section}

Now, using the previously obtained one-loop and three-loop doubled amplitudes, we derive the desired differential cross-sections $\mathrm{d}\sigma^{{\scriptscriptstyle(0)}}_{2\to 2}$ and $\mathrm{d}\sigma^{{\scriptscriptstyle(0)}}_{2\to 4}$ explicitly. We also compare them with the result obtained from squaring the amplitude.

We first identify the cuts to derive the differential cross-sections from the doubled amplitude. Since $D^{+-} = - \tilde{\Delta}^{+}$ and $D^{-+}=\tilde{\Delta}^{-}$, which define cuts \eqref{Cut_propagators}, then we evaluate the momentum integrals. Then the delta functions disappear, and we only have $D^{++}$ and $D^{--}$. Next, we use the momentum conservations appropriately, such as  $k_4 = -k_{123}$ and $k_{14} = -k_{23}$, to ensure that the propagators $D^{++}$ and $D^{--}$ carry the correct momentum labels.

Let us consider the one-loop double amplitude $\mathcal{A}^{{\scriptscriptstyle(1)}}_{1^+2^-3^-4^+}$. Substituting the $\tilde{\Delta}^{\pm}$, we have the relation between the one-loop doubled amplitude and the differential cross section $\mathrm{d}\sigma^{{\scriptscriptstyle(0)}}_{2\to2}$
\begin{equation}
\begin{aligned}
    \mathcal{A}^{{\scriptscriptstyle(1)}}_{1^+ 2^- 3^- 4^+} 
    &=
    \frac{\lambda^2}{2} \int \mathrm{d}\Pi_{p_{1}} \mathrm{d}\Pi_{p_{2}} \hat{\delta}(p_{1}+p_{2}-k_{1}-k_{4})\, 
    \\
    &= 
    \frac{1}{2} \int \mathrm{d}\Pi_{p_{1}} \mathrm{d}\Pi_{p_{2}} \hat{\delta}(p_{1}+p_{2}-k_{1}-k_{4})~ \big\lVert \mathcal{M}^{{\scriptscriptstyle(0)}}(k_1,k_4 \rightarrow p_{1},p_{2}) \big\lVert^{2}
    \,.
\end{aligned}
\end{equation}
where the Lorentz-invariant phase space measure $\mathrm{d}\Pi_{p}$
\begin{equation}
  \mathrm{d}\Pi_{p} = \frac{\mathrm{d}^{4}p}{(2\pi)^{4}} \hat{\delta}(p^{2}+m^{2}) \Theta(p^{0})\,.
\label{}\end{equation}
Here $\frac{1}{2}$ factor is the symmetric factor for two identical particles.

We now consider the three-loop doubled amplitude $\mathcal{A}^{{\scriptscriptstyle(3)}}_{1^+2^-3^-4^+}$. The relevant three-loop cut diagrams in a similar manner give us,
\begin{align}
\begin{split}
    \mathcal{A}^{{\scriptscriptstyle(3)}}_{1^+ 2^- 3^- 4^+} 
    & \sim 
    \frac{\lambda^4}{4!} \int \mathrm{d}\Pi_{k_{5}} \cdots\mathrm{d}\Pi_{k_{8}} \hat{\delta}(k_{5}+k_{6}+k_{7}+k_{8}-k_{1}-k_{4})
    \\&\qquad
    \times \Bigg(\tilde{D}^{+}_{145} + \tilde{D}^{+}_{146} + \tilde{D}^{+}_{147} + \tilde{D}^{+}_{148} + \tilde{D}^{+}_{156} 
    \\
    & \qquad \qquad 
    + \tilde{D}^{+}_{157} + \tilde{D}^{+}_{158} + \tilde{D}^{+}_{167} + \tilde{D}^{+}_{168} + D^{+}_{178} \Bigg)
    \\
    & \qquad 
    \times \Bigg(\tilde{D}^{-}_{235} + \tilde{D}^{-}_{236} + \tilde{D}^{-}_{237} + \tilde{D}^{-}_{238} + \tilde{D}^{-}_{256}
    \\
    & \qquad \qquad 
    + \tilde{D}^{-}_{257} + \tilde{D}^{-}_{258} + \tilde{D}^{-}_{267} + \tilde{D}^{-}_{268} + \tilde{D}^{-}_{278} \Bigg)
    \\
    & = \frac{1}{4!} \int \mathrm{d}\Pi_{k_{5}} \cdots\mathrm{d}\Pi_{k_{8}} \hat{\delta}(k_{5}+k_{6}+k_{7}+k_{8}-k_{1}-k_{4})
    \\
    & \qquad \qquad 
    \times \big\lVert \mathcal{M}^{(0)}(k_1,k_4 \rightarrow k_5,k_6,k_7,k_8) \big \Vert^{2} \,.
\end{split}
\end{align}
Again, $\frac{1}{4!}$ factor is the symmetric factor. 

Using the formalism we developed in the previous sections, we calculated the doubled amplitude $\mathcal{A}^{{\scriptscriptstyle(1)}}_{1^+ 2^- 3^- 4^+}$ and $\mathcal{A}^{{\scriptscriptstyle(3)}}_{1^+ 2^- 3^- 4^+}$ by solving the recursion relation and compared it with the differential cross-section, $\mathrm{d}\sigma^{{\scriptscriptstyle(0)}}_{2\to 2}$ and $\mathrm{d}\sigma^{{\scriptscriptstyle(0)}}_{2\to 4}$. We confirmed that these two quantities match exactly. This shows that this method can be used as an efficient approach for calculating cross-sections.

\section{Conclusion}\label{Sec:5}

In this work, we constructed a new method for calculating differential cross-sections based on the optical theorem and the quantum off-shell recursion. Our starting point was the derivation of an action formalism corresponding to the LTE, which we call the doubling prescription. This formalism is equivalent to the action of the Schwinger--Keldysh formalism by doubling the filed content. The scattering amplitudes in the doubling prescription automatically match the imaginary part of the amplitude in the original theory. The LTE also allows for the representation of this amplitude in terms of cut diagrams.

For efficient computation of doubled amplitudes, we derived the DS equation from the action in the doubling prescription. We also derived the perturbiner expansion, which is the generating function for the off-shell currents corresponding to doubled amplitudes, the key ingredient of the recursion relation for loop amplitudes. By substituting the perturbiner expansion into the DS equation, we obtained the quantum off-shell recursion. However, the off-shell current we obtained by solving the quantum off-shell recursion contains not only the doubled amplitude corresponding to our target but also unnecessary terms from different loop orders. To address this, we introduced and implemented an algorithm to efficiently filter out these unwanted terms.

Using this formalism, we directly computed the tree-level $ 2 \to 2 $ and $ 2 \to 4 $ differential cross sections in $ \phi^4 $ theory using the one-loop and three-loop doubled amplitudes, respectively. This approach avoids the conventional process of first computing the amplitude, squaring it, and then performing color and helicity sums. Instead, it employs the doubling prescription to calculate multi-loop amplitudes directly via recursion relations. While conventional methods are straightforward, they involve $ \mathcal{O}(N^2) $ complexity due to the squaring process and require extensive sums over colors and helicities, leading to computational overhead. In contrast, our method integrates color and helicity sums naturally within the multi-loop framework and circumvents the squaring.

To further develop this method as a practical tool for analyzing particle accelerator data, particularly from the LHC, it is essential to extend the approach to theories to gauge theories with color dynamics, such as QCD. Applying a method for phase space integration to compute total cross-sections will be an interesting future direction.


\section*{Acknowledgments}
We thank Poul Damgaard, Heribertus Bayu Hartanto, Seok Kim and Myeonghun Park for their useful comments and discussion. 
The work of VG has been supported in part by the INSPIRE-SHE scholarship awarded by the Department of Science and Technology, India.
The work of HL is supported in part by the National Research Foundation of Korea (NRF) grants, NRF-2023-K2A9A1A0609593811 and NRF RS-2024-0035-1197. 
The work of KL has been supported in part by appointment to the JRG Program at the APCTP through the Science and Technology Promotion Fund and Lottery Fund of the Korean Government. KL is also supported by the National Research Foundation of Korea(NRF) grant funded by the Korean government(MSIT) RS-2023-00249451 and the Korean Local Governments of Gyeongsangbuk-do Province and Pohang City.


\newpage	
\appendix


\section{Largest Time Equation}\label{App:A}
The unitarity of a quantum system can be represented in terms of the S-matrix concisely,
\begin{equation}
  \mathcal{S}^{\dagger} \mathcal{S} = \mathbf{1}\,,
\label{unitarity_S}\end{equation}
or in terms of the transfer matrix $\mathcal{T}$ defined as $\mathcal{S} = \mathbf{1} + i \mathcal{T}$, 
\begin{equation}
  - i \left(\mathcal{T}-\mathcal{T}^{\dagger}\right) = \mathcal{T}^{\dagger} \mathcal{T}\,.
\label{unitarity_T}\end{equation}
Note that the matrix elements of the S-matrix are represented by scattering amplitudes,
\begin{equation}
  \langle f|\mathcal{T}| i\rangle= \hat{\delta}^{4}(p_{i}-p_{f}) \mathcal{M}(i \to f)\,,
\label{}\end{equation}
where $ \hat{\delta}^4(k) = (2\pi)^{4}\delta^{4} (k)$. 

Substituting the completeness relation into the right-hand side of \eqref{unitarity_T} we have, 
\begin{equation}
\begin{aligned}
  2 \operatorname{Im} \mathcal{M}( i \to f) = \sum_{X} \int \mathrm{d} \Pi_{X}\,\mathcal{M}^{*}(f\to X) \mathcal{M}(i \to X) \hat{\delta}(k_{i} - k_{X})\,,
\end{aligned}\label{optical_thm}
\end{equation}
where on the right-hand side above we have a summation over all possible intermediate on-shell states $X$, and $\displaystyle\mathrm{d} \Pi_{X}$ is the phase-space integral,
\begin{equation}
  \mathrm{d} \Pi_{X} = \prod_{i\in X} \frac{\mathrm{d}^3 \vec{q}_{i}}{(2\pi)^{3}} \frac{1}{2E_{i}}\,.
\label{PhaseSpace}\end{equation}
The relation \eqref{optical_thm} is known as the optical theorem and it is a direct consequence of unitarity of the S-matrix. Assuming that we are studying $2\to2$ scattering where the initial and final states are identical, the optical theorem \eqref{optical_thm} reduces to the more familiar form,
\begin{equation}
  \operatorname{Im} \mathcal{M}\left(k_{1} k_{2} \to k_{1} k_{2}\right)=2 E_{\textrm{CM}} \big\lVert\vec{p}_{\textrm{CM}}\big\rVert \sum_{X}\sigma_{\mathrm{tot}}\left(k_{1} k_{2} \to X\right)\,,
\label{famousOpticalThm}\end{equation}
where $E_{\textrm{CM}}$ is the total center of mass energy and $\vec{p}_{\textrm{CM}}$ is the momentum of either of the two particles in the center of mass frame.

In perturbative QFT, the optical theorem can be implemented for each Feynman diagram that contributes to an amplitude according to the LSZ reduction formula. The Cutkosky rules prescribe the identification and evaluation of discontinuities (cuts) across diagrams, thereby enabling the computation of the imaginary parts of loop amplitudes directly \cite{Cutkosky:1960sp} since the imaginary part of a diagram is equivalent to the discontinuity. Cutkosky represented the discontinuity by imposing appropriate conditions on the cut lines, which meant replacing the cut internal propagators with the delta function,
\begin{equation}
  \frac{1}{p_{i}^{2}+m_{i}^{2}-i\epsilon} \to 2 \pi i \delta(p^{2}_{i}+m^{2}_{i})\,.
\label{CutkoskyReplacement}\end{equation}
%
So far we have discussed how unitarity can be manifested in individual Feynman diagrams contributing to the scattering amplitude via the Cutkosky rules. However, these rules rely on the explicit topology of each diagram, and illustrating each diagram would undermine the efficiency of the off-shell recursion framework. 

Thus, we instead focus on the largest time equation (LTE), which is more algebraic in nature and hence appropriate for the off-shell recursion framework. The LTE provides an alternative implication of unitarity for each individual diagram \cite{Veltman:1963th}. Instead of imposing the Cutkosky rules on each diagram, the LTE leads to a systematic representation of the optical theorem at the diagrammatic level.

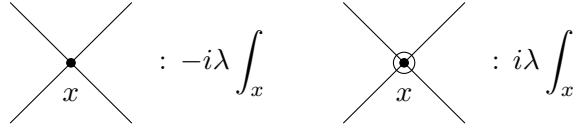
\begin{figure}[h]
\begin{center}
\begin{tikzpicture}[scale=0.8]
  	\draw (0,0) -- (2,2);
  	\filldraw[black] (1,1) circle (2pt) node[anchor=west]{\qquad~~: $\displaystyle -i \lambda \int_x$} node[below=6pt]{$x$};
  	\draw (2,0) -- (0,2);
\end{tikzpicture}
\qquad
\begin{tikzpicture}[scale=0.8]
  	\draw (0,0) -- (2,2);
  	\filldraw[black] (1,1) circle (2pt) node[anchor=west]{\qquad~~: $\displaystyle i \lambda \int_x$};
  	\draw (1,1) circle [radius=5pt]  node[below=6pt]{$x$};
  	\draw (2,0) -- (0,2);
\end{tikzpicture}
\end{center}
\vspace{-\baselineskip}
\caption{Interaction vertices}
\label{fig:InteractionVertices}
\end{figure}

\begin{figure}[h]
    \begin{center}
\begin{tikzpicture}[scale=0.8]
  	\draw (0,0) -- (2,0);
  	\filldraw[black] (0,0) circle (2pt) node[below=6pt]{$x$};
  	\filldraw[black] (2,0) circle (2pt) node[below=6pt]{$y$} node[anchor=west]{~~~ $D_{F}(x-y)$};
\end{tikzpicture}
\qquad
\begin{tikzpicture}[scale=0.8]
  	\draw (0,0) -- (2,0);
  	\filldraw[black] (0,0) circle (2pt) node[below=6pt]{$x$};
  	\filldraw[black] (2,0) circle (2pt) node[below=6pt]{$y$} node[anchor=west]{~~~ $\Delta^{+}(x-y)$};
  	\draw (0,0) circle [radius=5pt];
        \filldraw[black] (1.2,0) -- (0.9,-0.17) -- (0.9,0.17) -- (1.2,0);
\end{tikzpicture}
\\
\begin{tikzpicture}[scale=0.8]
  	\draw (0,0) -- (2,0);
  	\filldraw[black] (0,0) circle (2pt) node[below=6pt]{$x$};
  	\filldraw[black] (2,0) circle (2pt) node[below=6pt]{$y$} node[anchor=west]{~~~ $\Delta^{-}(x-y)$};
  	\draw (2,0) circle [radius=5pt];
        \filldraw[black] (0.8,0) -- (1.1,-0.17) -- (1.1,0.17) -- (0.8,0);
\end{tikzpicture}
\qquad
\begin{tikzpicture}[scale=0.8]
  	\draw (0,0) -- (2,0);
  	\filldraw[black] (0,0) circle (2pt) node[below=6pt]{$x$};
  	\filldraw[black] (2,0) circle (2pt) node[below=6pt]{$y$} node[anchor=west]{~~~ $D^{*}_{F}(x-y)$};
  	\draw (2,0) circle [radius=5pt];
 	\draw (0,0) circle [radius=5pt];
\end{tikzpicture}
\end{center}
\vspace{-\baselineskip}
\caption{Various propagators}
\label{fig:ThePropagators}
\end{figure}
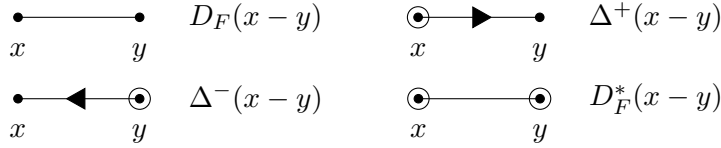

We demonstrate this using the scalar $\phi^4$ theory as our working example, and the arguments can be extended to more complicated theories. Let us consider a loop integrand $I_{\Gamma}(x_{1},x_{2}, \cdots x_{N})$ associated with a Feynman diagram $\Gamma$ with $N$-vertices, where $x_{i}$ are the positions of those interaction vertices. The key step for deriving the LTE is to enlarge the Feynman rules beyond the ones we have for the regular theory. The interaction vertices in the extension are given by the uncircled and the circled vertex as in figure \ref{fig:InteractionVertices}.
Thus, the uncircled vertex is identical to the one found in the regular theory, and the circled vertex is its complex conjugate. Furthermore, the new propagators, $D_{F}$, $\Delta^{\pm}$ and $D^{*}_{F}$, are given by all the two-point diagrams connecting the two kinds of vertices in figure \ref{fig:ThePropagators}.
And these propagators are explicitly defined as,
\begin{align}
  D_{F}(x-y) &=
  \Theta\left(x^{0}-y^{0}\right) \Delta^{+}(x-y)+\Theta\left(y^{0}-x^{0}\right) \Delta^{-}(x-y)\,,
  \\
  D_{F}^{*}(x-y) &=
  \Theta\left(y^{0}-x^{0}\right) \Delta^{+}(x-y)+\Theta\left(x^{0}-y^{0}\right) \Delta^{-}(x-y)\,,
\end{align}
where $\Theta\left(x^0 - y^0\right)$ is the Heaviside step function and we have,
\begin{equation}
  \Delta^{\pm}(x-y) 
  = 
  \int_k \, 2\pi \, e^{\pm ik \cdot (x-y)}\delta(k^2+m^2) \, \Theta(k^0) =\int \frac{\mathrm{d}^3\vec{k}}{(2 \pi)^{3} 2 E_{k}} e^{ \pm i k \cdot (x-y)}\,,
\label{}\end{equation}
and these are related by complex conjugation,
\begin{equation}
  \left(\Delta^{+}(x-y)\right)^{*}=\Delta^{+}(y-x) = \Delta^{-}(x-y)\,.
\label{}\end{equation}
Note that $D_F(x-y)$ is the usual Feynman propagator and the energy flows from the circled vertices to the uncircled vertices. This implies that diagrams with isolated vertices connected to only internal propagators give vanishing contributions as energy cannot flow from or to anywhere, for example,
\begin{align}\label{Isolated_circles_diagram}
\begin{tikzpicture}[anchor=base, baseline]
    \draw[thick] (-0.75,0) -- (-1,-0.5);
    \draw[thick] (-0.75,0) -- (-1,0.5);
    \draw[thick] (-0.75,0) to[out=60, in =180] (0,0.5);
    \draw[thick] (0,0.5) to[out=-150, in =150] (0,-0.5);
     \draw[thick] (0,0.5) to[out=-30, in =30] (0,-0.5);
    \draw[thick] (0.75,0) to[out=120, in =0] (0,0.5);
    \draw[thick] (0.75,0) to[out=-120, in =0] (0,-0.5);
    \draw[thick] (-0.75,0) to[out=-60, in =180] (0,-0.5);
    \draw[thick] (0.75,0) -- (1,-0.5);
    \draw[thick] (0.75,0) -- (1,0.5);
    \draw[fill=black] (-0.75,0) circle (0.05);
    \draw[fill=black] (0,0.5) circle (0.05);
    \draw (0,0.5) circle (0.125);
    \draw[fill=black] (0.75,0) circle (0.05);
    \draw[fill=black] (0,-0.5) circle (0.05);
    \node[] at (1,0) [right]{$=0$\,.};
\end{tikzpicture}
\end{align}
As energy flows only from the circled vertices to the uncircled vertices, from this point onwards we omit arrows in diagrams.

By enlarging the Feynman rules, we have $2^{N}$ duplicated diagrams associated with the initial diagram $\Gamma$ because there are now 2 vertex choices for each spacetime point $x_i$. The diagrams based on the enlarged Feynman rules are called \emph{cut diagrams} -- as the name suggests, we can assign the cutting rules. Without loss of generality, suppose that a vertex at $x_{k}$ has the largest time among all vertices, so $x^{0}_{k}> x_{i}^{0}$ for all $i \neq k$ and that it is connected to at least one other vertex. Then for all propagators involving $x_{k}$, we may replace $D_{F}$ (or $D^\ast_F$) with $\Delta^{\pm}$ (depending on the orientation). 

Using the properties of the enlarged Feynman rules, we have a crucial relation to a pair of cut diagrams. If we have two cut diagrams that are identical except that one has an uncircled vertex at $x_{k}$ while the other has a circled vertex at $x_{k}$, then it can be shown that the sum of these two cut diagrams vanishes. The relative sign difference arises because of the convention that a circled vertex carries a minus sign, whereas an uncircled vertex does not. Let $x_i$ be an arbitrary vertex adjacent to the largest time vertex at $x_{k}$. First, we assume that $x_{i}$ is an uncircled vertex, then we can show that the following relation, which is zoomed in a part of the pair of cut diagrams, always holds because of the sign difference between the uncircled and circled vertices: 
\begin{equation}
\begin{tikzpicture}[anchor=base, baseline, scale=0.8]
  	\draw (-0.7,0.7) -- (0,0);
  	\draw (-0.7,-0.7) -- (0,0);
	\draw[very thick] (0,0) -- (2,0); 
  	\draw (-1,0) -- (3,0) node[anchor=west]{\ $+$} node[below=20pt,pos=1/2]{\small $D_{F}(x_{i}-x_{k})\to \Delta^{-}(x_{i}-x_{k})$};
  	\draw (2.7,0.7) -- (2,0);
  	\draw (2.7,-0.7) -- (2,0);
  	\filldraw[black] (0,0) circle (2pt) node[below=6pt]{$x_{i}$} node[above=6pt]{\small $-i\lambda$};
  	\filldraw[black] (2,0) circle (2pt) node[below=6pt]{$x_{k}$} node[above=6pt]{\small $-i\lambda~~$};
\end{tikzpicture}
\begin{tikzpicture}[anchor=base, baseline, scale=0.8]
  	\draw (-0.7,0.7) -- (0,0);
  	\draw (-0.7,-0.7) -- (0,0);
  	\draw (-1,0) -- (3,0) node[anchor=west]{\ $= \,0\,.$} node[below=20pt,pos=1/2]{\small $\Delta^{-}(x_{i}-x_{k})$};
	\draw[very thick] (0,0) -- (2,0); 
  	\draw (2.7,0.7) -- (2,0);
  	\draw (2.7,-0.7) -- (2,0);
  	\filldraw[black] (0,0) circle (2pt) node[below=6pt]{$x_{i}$} node[above=6pt]{\small $-i\lambda$};
  	\filldraw[black] (2,0) circle (2pt) node[below=6pt]{$x_{k}$} node[above=6pt]{\small $i\lambda$};
  	\draw (2,0) circle [radius=5pt];
\end{tikzpicture}\label{}
\end{equation}
If we assume the alternative case where $x_{i}$ is a circled vertex, then the  relation still holds,
\begin{equation}
\begin{tikzpicture}[anchor=base, baseline, scale=0.8]
  	\draw (-0.7,0.7) -- (0,0);
  	\draw (-0.7,-0.7) -- (0,0);
	\draw[very thick] (0,0) -- (2,0); 
  	\draw (-1,0) -- (3,0) node[below=20pt,pos=1/2]{\small$\Delta^{+}(x_{i}-x_{k})$};
  	\draw (2.7,0.7) -- (2,0);
  	\draw (2.7,-0.7) -- (2,0);
  	\filldraw[black] (0,0) circle (2pt) node[below=6pt]{$x_{i}$} node[above=6pt]{\small $i\lambda$};
  	\filldraw[black] (2,0) circle (2pt) node[below=6pt]{$x_{k}$} node[above=6pt]{\small $-i\lambda~~$};
  	\draw (0,0) circle [radius=5pt];
\end{tikzpicture}
\begin{tikzpicture}[anchor=base, baseline, scale=0.8]
  	\draw (-0.7,0.7) -- (0,0);
  	\draw (-0.7,-0.7) -- (0,0);
  	\draw (-1,0) node[anchor=east]{$+\ $}-- (3,0) node[anchor=west]{\ $= \,0\,.$} node[below=20pt,pos=1/2]{\small  $D^{\ast}_{F}(x_{i}-x_{k})\to \Delta^{+}(x_{i}-x_{k})$};
	\draw[very thick] (0,0) -- (2,0); 
  	\draw (2.7,0.7) -- (2,0);
  	\draw (2.7,-0.7) -- (2,0);
  	\filldraw[black] (0,0) circle (2pt) node[below=6pt]{$x_{i}$} node[above=6pt]{\small $i\lambda$};
  	\filldraw[black] (2,0) circle (2pt) node[below=6pt]{$x_{k}$} node[above=6pt]{\small $i\lambda$};
  	\draw (0,0) circle [radius=5pt];
  	\draw (2,0) circle [radius=5pt];
\end{tikzpicture}
\end{equation}
Since this property holds for all adjacent vertices connected to the largest time vertex, and the fact that we could always choose a frame of reference to make any vertex $x_i$ have the largest time component, we can conclude that the sum of any two such cut diagrams that differ only in the circling of one vertex vanishes. 

Now consider the sum of all the $2^N$ diagrams associated with $\Gamma$, we can form $2^{\left(N-1\right)}$ pairs of diagrams such that they differ only in the circling of one arbitrarily chosen vertex. From the arguments shown above, it is obvious that the pairwise sum of all diagrams vanishes. If we denote the position of the circled vertex as $\bar{x}_{i}$ and reorganize the terms in the summation, then we have the following relation for the loop integrand $I_{\Gamma}$,
\begin{equation}
  I_{\Gamma}(x_{1},\cdots, x_{n}) + I_{\Gamma}(\bar{x}_{1},\cdots, \bar{x}_{n}) = - \sum_{\substack{\text{all possible circlings}\\\text{except fully circled}\\ \text{and fully uncircled}}} I_{\Gamma}(\cdots, x_{i},\cdots, \bar{x}_j, \cdots)\,.
\label{LTE}\end{equation}
This is called the LTE for a diagram $\Gamma$. Note that the left-hand side denotes the imaginary part of the diagram, and the right-hand side is the collection of all possible cut diagrams for a given Feynman diagram, which can be identified with the right-hand side of \eqref{optical_thm} through careful examination.

\subsection{Dyson--Schwinger equations for descendant fields}\label{App:B}

For the first descendant field, the analogous Dyson--Schwinger equations are,
\begin{equation}	
\begin{aligned}
 \psi^{AA'}_{x,z} 
  &= 
  D^{A+}_{xz}\eta^{+A'} + D^{A-}_{xz}\eta^{-A'}
  \\
  &\quad- \frac{\lambda^+}{3!}\int_y D^{A+}_{xy}\left(3\left(\varphi^+_y\right)^2\psi^{+A'}_{yz} + 3\frac{\hbar}{i}\varphi^+_y\psi^{++A'}_{yyz} + 3\frac{\hbar}{i}\psi^{+A'}_{yz}\psi^{++}_{yy} - \hbar^2\psi^{+++A'}_{yyyz}\right)
  \\
  &\quad+ \frac{\lambda^-}{3!}\int_y D^{A-}_{xy}\left(3\left(\varphi^-_y\right)^2\psi^{-A'}_{yz} + 3\frac{\hbar}{i}\varphi^-_y\psi^{--A'}_{yyz} + 3\frac{\hbar}{i}\psi^{-A'}_{yz}\psi^{--}_{yy} - \hbar^2\psi^{---A'}_{yyyz}\right)\,.
\end{aligned}\label{1stDescDysonSchwinger}
\end{equation}
Similarly for the second and third descendant fields we have the following descendant equations,
\begin{equation}	
\begin{aligned}
	\psi^{ABC}_{xzw} 
	&= -\frac{\lambda^+}{2!}\int_y D^{A,+}_{xy}\left(\vphantom{\frac{\hbar}{i}}\left(\varphi^+_y\right)^2\psi^{+BC}_{yzw} + 2\varphi^+_y\psi^{+B}_{yz}\psi^{+C}_{yw}+\frac{\hbar}{i}\varphi^+_y\psi^{++BC}_{y,y,z,w} 
	\right.\\&\qquad\qquad\qquad\qquad\left.
	+ \frac{\hbar}{i}\psi^{+C}_{yw}\psi^{++B}_{yyz} + \frac{\hbar}{i}\psi^{+BC}_{yzw}\psi^{++}_{yy} + \frac{\hbar}{i}\psi^{+B}_{yz}\psi^{++C}_{yyw}\right)
	\\
	&\quad+\frac{\lambda^-}{2!} \int_y D^{A-}_{xy}\left(\vphantom{\frac{\hbar}{i}}\left(\varphi^-_y\right)^2\psi^{-BC}_{yzw} + 2\varphi^-_y\psi^{-B}_{yz}\psi^{-C}_{yw}+ \frac{\hbar}{i}\varphi^-_y\psi^{--BC}_{yyzw} 
	\right.\\&\qquad\qquad\qquad\qquad\left.
	+ \frac{\hbar}{i}\psi^{-C}_{yw}\psi^{--B}_{yyz} + \frac{\hbar}{i}\psi^{-BC}_{y,z,w}\psi^{--}_{yy} + \frac{\hbar}{i}\psi^{-B}_{yz}\psi^{--C}_{yyw}\right)\vphantom{\int_y}
	\,,
\end{aligned}\label{2ndDesDysonSchwinger}
\end{equation}
and
\begin{equation}
\begin{aligned}
  \psi^{ABCD}_{xzwv} &= -\frac{\lambda^+}{2!}\int_y D^{A+}_{xy}\left(\left(\varphi^+_y\right)^2\psi^{+BCD}_{yzwv} + 2\psi^{+B}_{yz}\psi^{+C}_{yw}\psi^{+D}_{yv}+ 2\varphi^+_y\psi^{+B}_{yz}\psi^{+CD}_{ywv}
  \right.\\ 
	&\qquad\qquad\qquad\qquad\left.+ 2\varphi^+_y\psi^{+C}_{yw}\psi^{+BD}_{yzv}+ 2\varphi^+_y\psi^{+D}_{yv}\psi^{+BC}_{yzw}\right)
  \\
  &\quad +\frac{\lambda^-}{2!}\int_y D^{A-}_{xy}\left(\left(\varphi^-_y\right)^2\psi^{-BCD}_{yzwv} + 2\psi^{-B}_{yz}\psi^{-C}_{yw}\psi^{-D}_{yv}+ 2\varphi^-_y\psi^{-B}_{y,z}\psi^{-CD}_{ywv}
  \right.\\ 
  &\qquad\qquad\qquad\qquad\left.+ 2\varphi^-_y\psi^{-C}_{y,w}\psi^{-BD}_{yzv}+ 2\varphi^-_y\psi^{-D}_{yv}\psi^{-BC}_{yzw} \right)\,.
\end{aligned}\label{3rdDesDysonSchwinger}
\end{equation}

\newpage
\bibliography{reference}

\end{document}